\documentclass[runningheads]{svmult} 
 
\usepackage{makeidx}   
\usepackage{graphicx}  
\usepackage{subeqnar}  
\usepackage{multicol}  
\usepackage{physprbb}  
\makeindex             
 
\newcommand{\lsim}{\mathrel{\rlap{\lower4pt\hbox{\hskip0pt$\sim$}} 
\raise1pt\hbox{$<$}}}           
\newcommand{\gsim}{\mathrel{\rlap{\lower4pt\hbox{\hskip0pt$\sim$}} 
\raise1pt\hbox{$>$}}}           
\newcommand{\sfrac}[2]{\mbox{\footnotesize $\frac{#1}{#2}$}} 
 
\begin{document} 
\title*{Unifying aspects of light- and heavy-systems} 
\toctitle{Unifying aspects of light- and heavy-systems} 
%
%
\titlerunning{Unifying aspects of light- and heavy-systems} 
%
\author{Craig D.\ Roberts \inst{1,2} } 
\authorrunning{Craig D.\ Roberts} 
%
%
\institute{Physics Division, Bldg 203, Argonne National Laboratory\\ 
Argonne, Illinois 60439-4843, USA 
\and Fachbereich Physik, Universit\"at Rostock, D-18051 Rostock, 
Germany} 
 
\maketitle              
 
\begin{abstract} 
Dyson-Schwinger equations furnish a Poincar\'e covariant framework within which 
to study hadrons.  A particular feature is the existence of a nonperturbative, 
symmetry preserving truncation that enables the proof of exact results.  Key to 
the DSE's efficacious application is their expression of the materially 
important momentum-dependent dressing of parton propagators at infrared 
length-scales, which is responsible for the magnitude of constituent-quark 
masses and the length-scale characterising confinement in bound states.  A 
unified quantitative description of light- and heavy-quark systems is achieved 
by capitalising on these features. 
%
%
\end{abstract} 
 
\section{Introduction} 
This contribution provides an overview of one particular means by which a 
quantitative and intuitive understanding of strong interaction phenomena can 
be attained.  The broad framework is that of continuum strong QCD, by which I 
mean the continuum nonperturbative methods and models that can address these 
phenomena, especially those where a direct connection with QCD can be 
established, in one true limit or another.  Naturally, everyone has a 
favourite tool and, in this connection, the Dyson-Schwinger equations (DSEs) 
are mine \cite{cdragw}.  The framework is appropriate here because the last 
decade has seen a renaissance in its phenomenological application, with 
studies of phenomena as apparently unconnected as low-energy $\pi \pi$ 
scattering, $B \to D^\ast$ decays and the equation of state for a quark gluon 
plasma \cite{bastirev,reinhardrev,pieterrev}.  Indeed, the DSEs promise a 
single structure applicable to the gamut of strong interaction observables. 
 
Dyson-Schwinger equations provide a nonperturbative means of analysing a 
quantum field theory.  Derived from a theory's Euclidean space generating 
functional, they are an enumerable infinity of coupled integral equations 
whose solutions are the $n$-point Schwinger functions (Euclidean Green 
functions), which are the same matrix elements estimated in numerical 
simulations of lattice-QCD.  In theories with elementary fermions, the 
simplest of the DSEs is the {\it gap} equation, which is basic to studying 
dynamical symmetry breaking in systems as disparate as ferromagnets, 
superconductors and QCD.  The gap equation is a good example because it is 
familiar and has all the properties that characterise each DSE.  Its solution 
is a $2$-point function (the fermion propagator) but its kernel involves 
higher $n$-point functions; e.g., in a gauge theory, the kernel is 
constructed from the gauge-boson $2$-point function and fermion--gauge-boson 
vertex, a $3$-point function.  In addition, while a weak-coupling expansion 
yields all the diagrams of perturbation theory, a self-consistent solution of 
the gap equation exhibits nonperturbative effects unobtainable at any finite 
order in perturbation theory; e.g, dynamical chiral symmetry breaking (DCSB). 
 
The coupling between equations; namely, the fact that the equation for a 
given $m$-point function always involves at least one $n>m$-point function, 
necessitates a truncation of the tower of DSEs in order to define a tractable 
problem.  It is unsurprising that the best known truncation scheme is just 
the weak coupling expansion which reproduces every diagram in perturbation 
theory.  This scheme is systematic and valuable in the analysis of large 
momentum transfer processes because QCD is asymptotically free.  However, it 
precludes the study of nonperturbative effects, and hence something else is 
needed for the investigation of strongly interacting systems and bound state 
phenomena. 
 
In spite of the need for a truncation, gap equations have long been used 
effectively in obtaining nonperturbative information about many-body systems 
as, e.g., in the Nambu-Gorkov formalism for superconductivity.  The positive 
outcomes have been achieved through the simple expedient of employing the most 
rudimentary truncation, e.g., Hartree or Hartree-Fock, and comparing the 
results with observations.  Of course, agreement under these circumstances is 
not an unambiguous indication that the contributions omitted are small nor that 
the model expressed in the truncation is sound. However, it does justify 
further study, and an accumulation of good results is grounds for a concerted 
attempt to substantiate a reinterpretation of the truncation as the first term 
in a systematic and reliable approximation. 
 
The modern application of DSEs, notably, comparisons with and predictions of 
experimental data, can properly be said to rest on model assumptions. However, 
those assumptions can be tested within the framework and also via comparison 
with lattice-QCD simulations, and the predictions are excellent. Furthermore, 
progress in understanding the intimate connection between symmetries and 
truncation schemes has enabled exact results to be proved.  Herein I will 
briefly explain recent phenomenological applications and the foundation of 
their success, and focus especially on the links the approach provides between 
light- and heavy-quark phenomena.  It will become apparent that the 
momentum-dependent \textit{dressing} of the propagators of QCD's elementary 
excitations is a fundamental and observable feature of strong QCD. 
 
The article is organised as follows: 
Sec.\ \ref{sect2label} [p.\ \pageref{sect2label}] -- a review of DSE 
quiddities, especially in connection with the development of a nonperturbative, 
systematic and symmetry preserving truncation scheme, and the model-independent 
results whose proof its existence enables; 
Sec.\ \ref{sect4label} [p.\ \pageref{sect4label}] -- an illustration of the 
efficacious application of DSE methods to light-meson systems and the 
connections that may be made with the results of lattice-QCD simulations; 
Sec.\ \ref{sec:HQ} [p.\ \pageref{sec:HQ}] -- the natural extension of these 
methods to heavy-quark systems, with an explanation of the origin and 
derivation of heavy-quark symmetry limits and their confrontation with the 
real-world of finite quark masses; 
and Sec.\ \ref{sec:epilogue} [p.\ \pageref{sec:epilogue}] -- an epilogue. 
 
\section{Dyson-Schwinger Equations} 
\label{sect2label} 
\subsection{Gap Equation} 
The simplest DSE is the \textit{gap} equation, which describes how the 
propagation of a fermion is modified by its interactions with the medium being 
traversed.  In QCD that equation assumes the form:\footnote{A Euclidean metric 
is employed throughout, wherewith the scalar product of two four vectors is 
\mbox{$a\cdot b=\sum_{i=1}^4 a_i b_i$}; and I employ Hermitian Dirac-$\gamma$ 
matrices that obey \mbox{$\{\gamma_\mu,\gamma_\nu\} = 2\delta_{\mu\nu}$} and 
${\rm tr}\, \gamma_5\gamma_\mu\gamma_\nu\gamma_\rho\gamma_\sigma = 
-4\,\epsilon_{\mu\nu\rho\sigma}$, $\epsilon_{1234}=1$.} 
\begin{equation} 
S^{-1}(p)\!=\!Z_2(\zeta,\Lambda) \, i\, \gamma \cdot p 
    + Z_4(\zeta,\Lambda)\, m(\zeta) 
 + \Sigma'(p,\Lambda), 
    \label{gendse} 
\end{equation} 
wherein the dressed-quark self-energy is 
\begin{equation} 
\Sigma'(p,\Lambda) = Z_1(\zeta,\Lambda) \! 
 \int^\Lambda_q \!\!\! g^2D_{\mu\nu}(p-q) \, 
 \frac{\lambda^i}{2}\gamma_\mu \, 
   S(q) \, \Gamma^i_\nu(q,p). 
  \label{quarkSelf} 
\end{equation} 
Equations (\ref{gendse}), (\ref{quarkSelf}) constitute the renormalised DSE 
for the dressed-quark propagator.  In Eq.\ (\ref{quarkSelf}), $D_{\mu\nu}(k)$ 
is the renormalised dressed-gluon propagator, $\Gamma^a_\nu(q;p)$ is the 
renormalised dressed-quark-gluon vertex and $\int^\Lambda_q := \int^\Lambda 
d^4 q/(2\pi)^4$ represents a \textit{translationally-invariant} 
regularisation of the integral, with $\Lambda$ the regularisation 
mass-scale.\footnote{Only with a translationally invariant regularisation 
scheme can Ward-Takahashi identities be preserved, something that is crucial 
to ensuring vector and axial-vector current conservation.  The final stage of 
any calculation is to take the limit $\Lambda \to \infty$.} \ In addition, 
$Z_1(\zeta,\Lambda)$, $Z_2(\zeta,\Lambda)$ and $Z_4(\zeta,\Lambda)$ are, 
respectively, Lagrangian renormalisation constants for the quark-gluon 
vertex, quark wave function and quark mass-term, which depend on the 
renormalisation point, $\zeta$, and the regularisation mass-scale, as does 
the gauge-independent mass renormalisation constant, 
\begin{equation} 
\label{Zmass} Z_m(\zeta^2,\Lambda^2) = Z_4(\zeta^2,\Lambda^2) \, 
Z_2^{-1}(\zeta^2,\Lambda^2)\, , 
\end{equation} 
whereby the renormalised running-mass is related to the bare mass: 
\begin{equation} 
\label{mzeta}  m(\zeta) = Z_m^{-1}(\zeta^2,\Lambda^2) \,m_{\rm bm}(\Lambda) \,. 
\end{equation} 
When $\zeta$ is very large the running-mass can be evaluated in perturbation 
theory, which gives 
\begin{equation} 
\label{hatm} 
m(\zeta) = 
\frac{\hat m}{(\ln \zeta/\Lambda_{\rm QCD})^{\gamma_m}\!\!\!\!\!\!\!} 
\;\;\;\;\; , \; \gamma_m = 12/(33 - 2 N_f)\,. 
\end{equation} 
Here $N_f$ is the number of current-quark flavours that contribute actively 
to the running coupling, and $\Lambda_{\rm QCD}$ and $\hat m$ are 
renormalisation group invariants. 
 
The solution of Eq.~(\ref{gendse}) is the dressed-quark propagator and takes 
the form 
\begin{equation} 
 S^{-1} (p) =  i \gamma\cdot p \, A(p^2,\zeta^2) + B(p^2,\zeta^2) \\ 
 \equiv \frac{1}{Z(p^2,\zeta^2)}\left[ i\gamma\cdot p + M(p^2)\right]. 
\label{sinvp} 
\end{equation} 
It is obtained by solving the gap equation subject to the renormalisation 
condition that at some large spacelike $\zeta^2$ 
\begin{equation} 
\label{renormS} \left.S^{-1}(p)\right|_{p^2=\zeta^2} = i\gamma\cdot p + 
m(\zeta)\,. 
\end{equation} 
 
The observations made in in the Introduction are now manifest.  The gap 
equation is a nonlinear integral equation for $S(p)$ and can therefore yield 
much-needed nonperturbative information.  However, the kernel involves the 
two-point function $D_{\mu\nu}(k)$ and the three-point function 
$\Gamma^a_\nu(q;p)$.  The equation is consequently coupled to the DSEs these 
functions satisfy and hence a manageable problem is obtained only once a 
truncation scheme is specified. 
 
\subsection{Nonperturbative Truncation} 
\label{subsec:truncation} 
To understand why Eq.\ (\ref{gendse}) is called a gap equation, consider the 
chiral limit, which is readily defined \cite{mrt98} because QCD exhibits 
asymptotic freedom and implemented in the gap equation by employing 
\cite{mr97} 
\begin{equation} 
\label{chirallimit} Z_2(\zeta^2,\Lambda^2) \, 
m_{\rm bm}(\Lambda) \equiv 0\,, \;\; \Lambda \gg \zeta\,. 
\end{equation} 
It is noteworthy that for finite $\zeta$ and $\Lambda \to \infty$, the left 
hand side (l.h.s.) of Eq.\ (\ref{chirallimit}) is identically zero, by 
definition, because the mass term in QCD's Lagrangian density is 
renormalisation-point-independent.  The condition specified in Eq.\ 
(\ref{chirallimit}), on the other hand, effects the result that at the 
(perturbative) renormalisation point there is no mass-scale associated with 
explicit chiral symmetry breaking, which is the essence of the chiral limit. 
An equivalent statement is that one obtains the chiral limit when the 
renorma\-li\-sa\-tion-point-invariant current-quark mass vanishes; namely, 
$\hat m = 0$ in Eq.\ (\ref{hatm}).  In this case the theory is chirally 
symmetric, and a perturbative evaluation of the dressed-quark propagator from 
Eq.\ (\ref{gendse}) gives 
\begin{equation} 
\label{Bpert0} 
B_{\rm pert}^0(p^2) := \lim_{m\to 0} B_{\rm pert}(p^2) = \lim_{m\to 0} m 
\left( 1 - \frac{\alpha}{\pi} \ln \left[\frac{p^2}{m^2} \right] + \ldots 
\right) \equiv 0\,; 
\end{equation} 
viz., the perturbative mass function is identically zero in the chiral limit. 
It follows that there is no gap between the top level in the quark's filled 
negative-energy Dirac sea and the lowest positive energy level. 
 
However, suppose one has at hand a truncation scheme other than perturbation 
theory and that subject to this scheme Eq.\ (\ref{gendse}) possessed a chiral 
limit solution $B^0(p^2)\not\equiv 0$.  Then interactions between the quark 
and the virtual quanta populating the ground state would have 
nonperturbatively generated a mass gap.  The appearance of such a gap breaks 
the theory's chiral symmetry.  This shows that the gap equation can be an 
important tool for studying DCSB, and it has long been used to explore this 
phenomenon in both QED and QCD \cite{cdragw}. 
 
The gap equation's kernel is formed from a product of the dressed-gluon 
propagator and dres\-sed-quark-gluon vertex but in proposing and developing a 
truncation scheme it is insufficient to focus only on this kernel 
\cite{mr97,raya}.  The gap equation can only be a useful tool for studying 
DCSB if the truncation itself does not destroy chiral symmetry. 
 
Chiral symmetry is expressed via the axial-vector Ward-Takahashi 
identity:\label{chiralsymmetry} 
\begin{equation} 
P_\mu \, \Gamma_{5\mu}(k;P)  =  S^{-1}(k_+)\, i\gamma_5 + i\gamma_5 
\,S^{-1}(k_-)\,,\; k_\pm = k\pm  P/2 , \label{avwti} 
\end{equation} 
wherein $\Gamma_{5\mu}(k;P)$ is the dressed axial-vector vertex.  This 
three-point function satisfies an inhomogeneous Bethe-Salpeter equation 
(BSE): 
\begin{equation} 
\label{avbse} 
\left[\Gamma_{5\mu}(k;P)\right]_{tu} 
 =  Z_2 \left[\gamma_5\gamma_\mu\right]_{tu} + \int^\Lambda_q 
[S(q_+) \Gamma_{5\mu}(q;P) S(q_-)]_{sr} K_{tu}^{rs}(q,k;P)\,, 
\end{equation} 
in which $K(q,k;P)$ is the fully-amputated quark-antiquark scattering kernel, 
and the colour-, Dirac- and flavour-matrix structure of the elements in the 
equation is denoted by the indices $r,s,t,u$.  The Ward-Takahashi identity, 
Eq.~(\ref{avwti}), entails that an intimate relation exists between the 
kernel in the gap equation and that in the BSE. (This is another example of 
the coupling between DSEs.) Therefore an understanding of chiral symmetry and 
its dynamical breaking can only be obtained with a nonperturbative truncation 
scheme that preserves this relation, and hence guarantees Eq.~(\ref{avwti}) 
without a \textit{fine-tuning} of model-dependent parameters. 
 
\subsubsection{Rainbow-ladder truncation.} 
At least one such scheme exists \cite{truncscheme}.  Its leading-order term 
is the so-called re\-nor\-ma\-li\-sa\-tion-group-improved rainbow-ladder 
truncation, whose analogue in the many body problem is an Hartree-Fock 
truncation of the one-body (Dyson) equation combined with a consistent 
ladder-truncation of the related two-body (Bethe-Salpeter) equation.  To 
understand the origin of this leading-order term, observe that the 
dressed-ladder truncation of the quark-antiquark scattering kernel is 
expressed in Eq.\ (\ref{avbse}) via 
\begin{eqnarray} 
\nonumber \lefteqn{ [ L(q,k;P)]_{t u}^{t^\prime u^\prime } \, 
[\Gamma_{5\mu}(q;P)]_{u^\prime t^\prime }  :=[S(q_+) \Gamma_{5\mu}(q;P) 
S(q_-)]_{sr}\, K_{tu}^{rs}(q,k;P)}\\ 
\nonumber & =& - g^2(\zeta^2)\, D_{\rho\sigma}(k-q) \, \\ 
& & \times \left[\rule{0mm}{0.7\baselineskip} 
\Gamma^a_\rho(k_+,q_+)\,S(q_+) \right]_{tt^\prime} 
\left[\rule{0mm}{0.7\baselineskip} S(q_-)\,\Gamma^a_\sigma(q_-,k_-) 
\right]_{u^\prime u}\, [\Gamma_{5\mu}(q;P)]_{t^\prime u^\prime} \label{RL1} 
\end{eqnarray} 
wherein I have only made explicit the renormalisation point dependence of the 
coupling.  One can exploit multiplicative renormalisability and asymptotic 
freedom to demonstrate that on the kinematic domain for which $Q^2:=(k-q)^2 
\sim k^2\sim q^2$ is large and spacelike 
\begin{equation} 
\label{LqkPuv} 
[L(q,k;P)]_{t u}^{t^\prime u^\prime } 
= - 
4 \pi \alpha(Q^2) \, D_{\rho\sigma}^{\rm free}(Q)\, 
\left[\rule{0mm}{0.7\baselineskip} 
        \frac{\lambda^a}{2}\gamma_\rho \,S^{\rm free}(q_+)\right]_{tt^\prime} 
\left[\rule{0mm}{0.7\baselineskip}S^{\rm free}(q_-)\, 
        \frac{\lambda^a}{2}\gamma_\sigma\right]_{u^\prime u} \! , 
\end{equation} 
where $\alpha(Q^2)$ is the strong running coupling and, e.g., $S^{\rm free}$ 
is the free quark propagator.  It follows that on this domain the r.h.s.\ of 
Eq.\ (\ref{LqkPuv}) describes the leading contribution to the complete 
quark-antiquark scattering kernel, $K_{tu}^{rs}(q,k;P)$, with all other 
contributions suppressed by at least one additional power of $1/Q^2$. 
 
The renormalisation-group-improved ladder-truncation supposes that 
\begin{equation} 
\label{ladder} 
K_{tu}^{rs}(q,k;P) = - 4 \pi \,\alpha(Q^2) \, D_{\rho\sigma}^{\rm free}(Q)\, 
\left[\rule{0mm}{0.7\baselineskip} 
        \frac{\lambda^a}{2}\gamma_\rho \right]_{ts} 
\left[\rule{0mm}{0.7\baselineskip} 
        \frac{\lambda^a}{2}\gamma_\sigma\right]_{r u} 
\end{equation} 
is also a good approximation on the infrared domain and is thus an assumption 
about the long-range ($Q^2 \lsim 1\,$ GeV$^2$) behaviour of the interaction. 
Combining Eq.\ (\ref{ladder}) with the requirement that Eq.\ (\ref{avwti}) be 
automatically satisfied leads to the renormalisation-group-improved 
rainbow-truncation of the gap equation: 
\begin{equation} 
\label{rainbowdse} S^{-1}(p) = Z_2 \,(i\gamma\cdot p + m_{\rm bm}) + 
\int^\Lambda_q \! 4 \pi \,\alpha(Q^2) \, D_{\mu\nu}^{\rm free} (p-q) 
\frac{\lambda^a}{2}\gamma_\mu \, S(q) \, \frac{\lambda^a}{2} \gamma_\nu\,. 
\end{equation} 
This rainbow-ladder truncation provides the foundation for an explanation of 
a wide range of hadronic phenomena \cite{pieterrev}. 
 
\subsection{Systematic Procedure} 
\label{subsec:systematic} 
The truncation scheme of Ref.\ \cite{truncscheme} is a dressed-loop expansion 
of the dressed-quark-gluon vertices that appear in the half-amputated 
dressed-quark-anti\-quark scattering matrix: $S^2 K$, a 
re\-nor\-ma\-li\-sa\-tion-group in\-va\-ri\-ant \cite{detmold}.  All 
$n$-point functions involved thereafter in connecting two particular 
quark-gluon vertices are \textit{fully dressed}. The effect of this 
truncation in the gap equation, Eq.~(\ref{gendse}), is realised through the 
following representation of the dressed-quark-gluon vertex, $i \Gamma_\mu^a = 
\frac{i}{2}\lambda^a\,\Gamma_\mu = l^a \Gamma_\mu$: 
\begin{eqnarray} 
\nonumber 
\lefteqn{Z_1 \Gamma_\mu(k,p)   =   \gamma_\mu +  \frac{1}{2 N_c} 
\int_\ell^\Lambda\! g^2 D_{\rho\sigma}(p-\ell) 
\gamma_\rho S(\ell+k-p) \gamma_\mu S(\ell) 
\gamma_\sigma}\\ 
\nonumber &+ & \frac{N_c}{2}\int_\ell^\Lambda\! g^2\, 
D_{\sigma^\prime \sigma}(\ell) \, D_{\tau^\prime\tau}(\ell+k-p)\, 
\gamma_{\tau^\prime} \, S(p-\ell)\, 
\gamma_{\sigma^\prime}\, 
\Gamma^{3g}_{\sigma\tau\mu}(\ell,-k,k-p) + [\ldots]\,. \\ 
\label{vtxexpand} 
\end{eqnarray} 
Here $\Gamma^{3g}$ is the dressed-three-gluon vertex and it is readily 
apparent that the lowest order contribution to each term written explicitly 
is O$(g^2)$.  The ellipsis represents terms whose leading contribution is 
O$(g^4)$; viz., the crossed-box and two-rung dressed-gluon ladder diagrams, 
and also terms of higher leading-order. 
 
This expansion of $S^2 K$, with its implications for other $n$-point 
functions, yields an ordered truncation of the DSEs that guarantees, 
term-by-term, the preservation of vector and axial-vector Ward-Takahashi 
identities, a feature that has been exploited 
\cite{mrt98,marisAdelaide,mishaSVY} to establish exact results in QCD.  It is 
readily seen that inserting Eq.\ (\ref{vtxexpand}) into Eq.\ (\ref{gendse}) 
provides the rule by which the rainbow-ladder truncation can be 
systematically improved. 
 
\subsubsection{Planar vertex.} 
\begin{figure}[t] 
\centerline{\includegraphics[width=.83\textwidth]{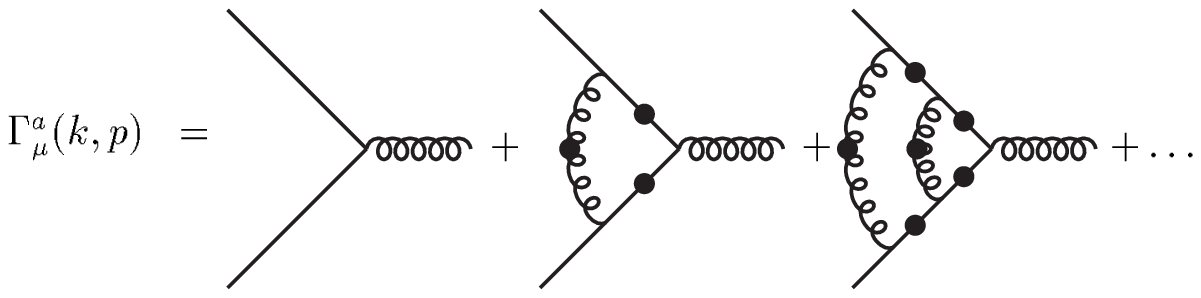}} 
\caption{\label{Gamma_inf} Integral equation for a planar dressed-quark-gluon 
vertex obtained by neglecting contributions associated with explicit gluon 
self-interactions.  Solid circles indicate fully dressed propagators.  The 
vertices are not dressed.  (Adapted from Ref.\ \protect\cite{detmold}.)} 
\end{figure}

The effect of the complete vertex in Eq.\ (\ref{vtxexpand}) on the solutions 
of the gap equation is unknown.  However, insights have been drawn from a 
study \cite{detmold} of a more modest problem obtained by retaining only the 
sum of dressed-gluon ladders; i.e., the vertex depicted in Fig.\ 
\ref{Gamma_inf}.  The elucidation is particularly transparent when one 
employs \cite{mn83} 
\begin{equation} 
\label{mnmodel} {\cal D}_{\mu\nu}(k):= g^2 \, D_{\mu\nu}(k) = 
\left(\delta_{\mu\nu} - \frac{k_\mu k_\nu}{k^2}\right) (2\pi)^4\, {\cal G}^2 
\, \delta^4(k) 
\end{equation} 
for the dressed-gluon line, which defines an ultraviolet finite model so that 
the regularisation mass-scale can be removed to infinity and the 
renormalisation constants set equal to one.\footnote{The constant ${\cal G}$ 
sets the model's mass-scale and using ${\cal G}=1$ simply means that all 
mass-dimensioned quantities are measured in units of ${\cal G}$.} \ This 
model has many positive features in common with the class of 
renormalisation-group-improved rainbow-ladder models and its particular 
momentum-dependence works to advantage in reducing integral equations to 
algebraic equations with similar qualitative features.  There is naturally a 
drawback: the simple momentum dependence also leads to some model-dependent 
artefacts, but they are easily identified and hence not cause for concern. 
 
The general form of the dressed-quark gluon vertex involves twelve distinct 
scalar form factors but using Eq.~(\ref{mnmodel}) only $\Gamma_\mu(p) 
:=\Gamma_\mu(p,p)$ contributes to the gap equation, which considerably 
simplifies the analysis.  The summation depicted in Fig.\ \ref{Gamma_inf} is 
expressed via 
\begin{equation} 
\label{vtxalgebraic} \Gamma_\mu(p) = \gamma_\mu + \frac{1}{8}\,\gamma_\rho\, 
S(p)\, \Gamma_\mu(p)\, S(p)\, \gamma_\rho\,, 
\end{equation} 
which supports a solution 
\begin{equation} 
\label{vtxinfty} \Gamma_\mu(p) = \alpha_1(p^2)\, \gamma_\mu + \, 
\alpha_2(p^2)\, \gamma\cdot p\,p_\mu - \, \alpha_3(p^2)\, i \,p_\mu\,. 
\end{equation} 
 
One can re-express this vertex as \begin{eqnarray} 
\Gamma_\mu(p) = \sum_{i=0}^\infty\,\Gamma_\mu^i(p) = 
\sum_{i=0}^\infty\, \left[ \alpha^i_1(p^2)\, \gamma_\mu + \, 
\alpha^i_2(p^2)\, \gamma\cdot p\,p_\mu - \, \alpha^i_3(p^2)\, i 
\,p_\mu\right], \label{vtxi} 
\end{eqnarray} 
where the superscript enumerates the order of the iterate: $\Gamma_\mu^{i=0}$ 
is the bare vertex, 
\begin{equation} 
\label{alpha0} 
\begin{array}{cc} 
\alpha_1^0 = 1\,,\; & \alpha_2^0=0=\alpha_3^0\,; 
\end{array} 
\end{equation} 
$\Gamma_\mu^{i=1}$ is the result of inserting this into the r.h.s.\ of 
Eq~(\ref{vtxalgebraic}) to obtain the one-rung dressed-gluon correction; 
$\Gamma_\mu^{i=2}$ is the result of inserting $\Gamma_\mu^{i=1}$, and is 
therefore the two-rung dressed-gluon correction; etc.  A key observation 
\cite{detmold} is that each iterate is related to its precursor via a simple 
recursion relation and, substituting Eq.~(\ref{vtxi}), that recursion yields 
($s=p^2$) 
\begin{equation} 
\label{matrixrecurs} \mbox{\boldmath $\alpha$}^{i+1}(s):= 
\left( 
\begin{array}{l} 
\rule{0ex}{2.5ex}\alpha_1^{i+1}(s) \\\rule{0ex}{2.5ex} \alpha_2^{i+1}(s) \\ 
\rule{0ex}{2.5ex}\alpha_3^{i+1}(s) 
\end{array}\right) 
= {\cal O}(s;A,B)\, \mbox{\boldmath $\alpha$}^{i}(s)\,, 
\end{equation} 
\begin{equation} 
{\cal O}(s;A,B) = \frac{1}{4} \,\frac{1}{\Delta^2} 
\left( 
\begin{array}{ccc} 
\rule{0ex}{2.5ex} - \Delta & 0 & 0\\ 
\rule{0ex}{2.5ex} 2  A^2 & s A^2 - B^2 & 2 A B \\ 
\rule{0ex}{2.5ex} 4 A B & 4 s A B & 2 (B^2 - s A^2) 
\end{array} \right), 
\end{equation} 
$\Delta = s A^2(s) + B^2(s)$.  It follows that 
\begin{equation} 
\label{boldalpha} 
\mbox{\boldmath $\alpha$} 
= \left(\sum_{i=1}^\infty\, {\cal O}^i\right) \, \mbox{\boldmath $\alpha$}^0 
= \frac{1}{1 - {\cal O}} \,\mbox{\boldmath $\alpha$}^0\, 
\end{equation} 
and hence, using Eq.\ (\ref{alpha0}), 
\begin{eqnarray} 
\nonumber 
\alpha_1 & = & \frac{4\, \Delta}{1 + 4\,\Delta}\,,\\ 
 \alpha_2 & = & \frac{- \,8 \,A^2} { 
1 + 2\,( B^2 - s\,A^2) - 8\,\Delta^2 } \, 
\frac{1 + 2\,\Delta}{1 + 4\,\Delta} \,, \label{alpharesults}\\ 
\nonumber \alpha_3 & = & 
\frac{- 8 \,A B}{ 1 + 2 (B^2 - s\, A^2) - 8\,\Delta^2 }\,. 
\end{eqnarray} 
 
The recursion relation thus leads to a closed form for the 
gluon-ladder-dressed quark-gluon vertex in Fig. \ref{Gamma_inf}; viz., Eqs.\ 
(\ref{vtxinfty}), (\ref{alpharesults}).  Its momentum-dependence is 
determined by that of the dressed-quark propagator, which is obtained by 
solving the gap equation, itself constructed with this vertex.  Using 
Eq.~(\ref{mnmodel}), that gap equation is 
\begin{equation} 
S^{-1}(p) = 
 i \gamma\cdot p + m + \gamma_\mu \, S(p) \,\Gamma_\mu(p) \\ 
\label{gapmodel} 
\end{equation} 
whereupon the substitution of Eq.~(\ref{vtxinfty}) gives 
\begin{eqnarray} 
A(s) & = & 1 +\frac{1}{s A^2 + B^2} \left[ A \, (2 \alpha_1 - s 
\alpha_2) - B \,\alpha_3\right]\,,\label{Afull}\\ 
B(s) & = & m+ \frac{1}{s A^2 + B^2} \left[ 
B \, ( 4 \alpha_1 + s \alpha_2) - s A \,\alpha_3 \right]\,. \label{Bfull} 
\end{eqnarray} 
Equations (\ref{Afull}), (\ref{Bfull}), completed using 
Eqs.~(\ref{alpharesults}), form a closed algebraic system.  It can easily be 
solved numerically, and that yields simultaneously the complete 
gluon-ladder-dressed vertex and the propagator for a quark fully dressed via 
gluons coupling through this nonperturbative vertex.  Furthermore, it is 
apparent that in the chiral limit, $m=0$, a realisation of chiral symmetry in 
the Wigner-Weyl mode, which is expressed via the $B\equiv 0$ solution of the 
gap equation, is always admissible.  This is the solution anticipated in Eq.\ 
(\ref{Bpert0}). 
 
\begin{figure}[t] 
\centerline{\includegraphics[width=.60\textwidth]{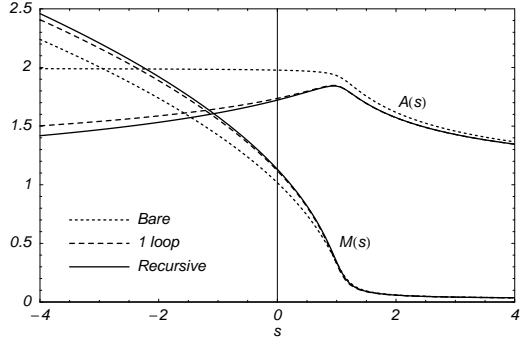}} 
\caption{\label{AM0plot} $A(s)$, $M(s)$ obtained from Eqs.\ 
(\protect\ref{alpharesults}), (\protect\ref{Afull}), (\protect\ref{Bfull}) 
with $m=0.023$ (\textit{solid line}).  All dimensioned quantities are 
expressed in units of ${\cal G}$ in Eq.~(\protect\ref{mnmodel}).  For 
comparison, the results obtained with the zeroth-order vertex (\textit{dotted 
line}) and the one-loop vertex (\textit{dashed line}) are also plotted. 
(Adapted from Ref.\ \protect\cite{detmold}.)} 
\end{figure} 
 
The chiral limit gap equation also admits a Nambu-Goldstone mode solution 
whose $p^2\simeq 0$ properties are unambiguously related to those of the 
$m\neq 0$ solution, a feature also evident in QCD \cite{mishaSVY}.  A 
complete solution of Eq.\ (\ref{gapmodel}) is available numerically, and 
results for the dressed-quark propagator are depicted in Fig.\ \ref{AM0plot}. 
It is readily seen that the complete resummation of dressed-gluon ladders 
gives a dressed-quark propagator that is little different from that obtained 
with the one-loop-corrected vertex; and there is no material difference from 
the result obtained using the zeroth-order vertex.  Similar observations 
apply to the vertex itself.  The scale of these modest effects can be 
quantified by a comparison between the values of $M(s=0)=B(0)/A(0)$ 
calculated using vertices dressed at different orders: 
\begin{equation} 
\begin{array}{l||c|c|c|c} 
\sum_{i=0,N}\Gamma_\mu^i & N=0 & N=1 & N=2 & N=\infty\\[1.5ex]\hline 
M(0) & \rule{0em}{2ex} 1 & 1.105 &  1.115 & 1.117 
\end{array} 
\end{equation} 
The rainbow truncation of the gap equation is accurate to within 12\% and 
adding just one gluon ladder gives 1\% accuracy.  It is important to couple 
this with an understanding of how the vertex resummation affects the 
Bethe-Salpeter kernel. 
 
\subsubsection{Vertex-consistent Bethe-Salpeter kernel.} 
\begin{figure}[t] 
\centerline{\includegraphics[width=.60\textwidth]{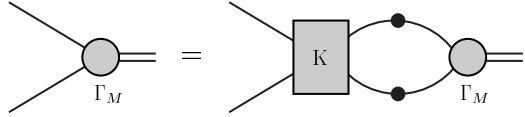}} 
\caption{\label{BSEpic} Homogeneous BSE, Eq.\ (\protect\ref{bsegen}).  Filled 
circles: dressed propagators or vertices; $K$ is the dressed-quark-antiquark 
scattering kernel.  A systematic truncation of $S^2 K$ is the key to 
preserving Ward-Takahashi identities \protect\cite{truncscheme,herman}. 
(Adapted from Ref.\ \protect\cite{detmold}.)} 
\end{figure}

The renormalised homogeneous BSE for the quark-antiquark channel denoted by 
$M$ can be expressed 
\begin{equation} 
\label{bsegen} 
    [\Gamma_M(k;P)]_{tu} =\int_q^\Lambda\! [\chi_M(q;P)]_{sr}\, 
    [ K(k,q;P)]_{tu}^{rs}\,, 
\end{equation} 
where: $\Gamma_M(k;P)$ is the meson's Bethe-Salpeter amplitude, $k$ is the 
relative momentum of the quark-antiquark pair, $P$ is their total momentum; and 
\begin{equation} 
\label{chiM} 
\chi_M(k;P) = S(k_+)\, \Gamma_M(k;P) \,S(k_-)\,. 
\end{equation} 
Equation (\ref{bsegen}), depicted in Fig.\ \ref{BSEpic}, describes the 
residue at a pole in the solution of an inhomogeneous BSE; e.g., the lowest 
mass pole solution of Eq.\ (\ref{avbse}) is identified with the 
pion.\footnote{The canonical normalisation of a Bethe-Salpeter amplitude is 
fixed by requiring that the bound state contribute with unit residue to the 
fully-amputated quark-antiquark scattering amplitude: $M = K + K (SS) K + 
[\ldots]$.  See, e.g., Ref.\ \protect\cite{llewellyn}.} 
 
I noted on p.\ \pageref{chiralsymmetry} that the automatic preservation of 
Ward-Takahashi identities in those channels related to strong interaction 
observables requires a conspiracy between the dressed-quark-gluon vertex and 
the Bethe-Salpeter kernel \cite{truncscheme,herman}.  A systematic procedure 
for building that kernel follows \cite{detmold} from the observation 
\cite{herman} that the gap equation can be expressed via 
\begin{equation} 
\frac{\delta \Gamma[S] }{\delta S} = 0 \,, 
\end{equation} 
where $\Gamma[S]$ is a Cornwall-Jackiw-Tomboulis-like effective action 
\cite{haymaker}.  The Bethe-Salpeter kernel is then obtained via an 
additional functional derivative: 
\begin{equation} 
K_{tu}^{rs} = - \frac{\delta \Sigma_{tu}}{\delta S_{rs}}\,. \label{KCJT} 
\end{equation} 
 
With the vertex depicted in Fig.\ \ref{Gamma_inf}, the $n$-th order 
contribution to the kernel is obtained from the $n$-loop contribution to the 
self energy: 
\begin{equation} 
    \Sigma^n(p)= - \int_q^\Lambda\! {\cal D}_{\mu\nu}(p-q) \, l^a \gamma_\mu \, 
    S(q) l^a \,\Gamma_{\nu}^n(q,p). 
\end{equation} 
Since $\Gamma_\mu(p,q)$ itself depends on $S$ then Eq.~(\ref{KCJT}) yields the 
Bethe-Salpeter kernel as a sum of two terms and hence Eq.\ (\ref{bsegen}) 
assumes the form 
\begin{equation} 
\Gamma_M(k;P)  =  \int_q^\Lambda 
{\cal D}_{\mu\nu}(k - q)\,l^a \gamma_\mu \left[\rule{0em}{3ex} \chi_M(q;P) \, 
l^a\, 
\Gamma_\nu(q_-,k_-) + S(q_+) \, \Lambda_{M \nu}^{a}(q,k;P)\right], 
\label{genbsenL1} 
\end{equation} 
where I have used the mnemonic 
\begin{equation} 
\Lambda_{M \nu}^{a}(q,k;P) = \sum_{n=0}^\infty \Lambda_{M \nu}^{a;n}(q,k;P)\,. 
\label{Lambdatotal} 
\end{equation} 
 
\begin{figure}[t] 
\centerline{\includegraphics[width=.66\textwidth]{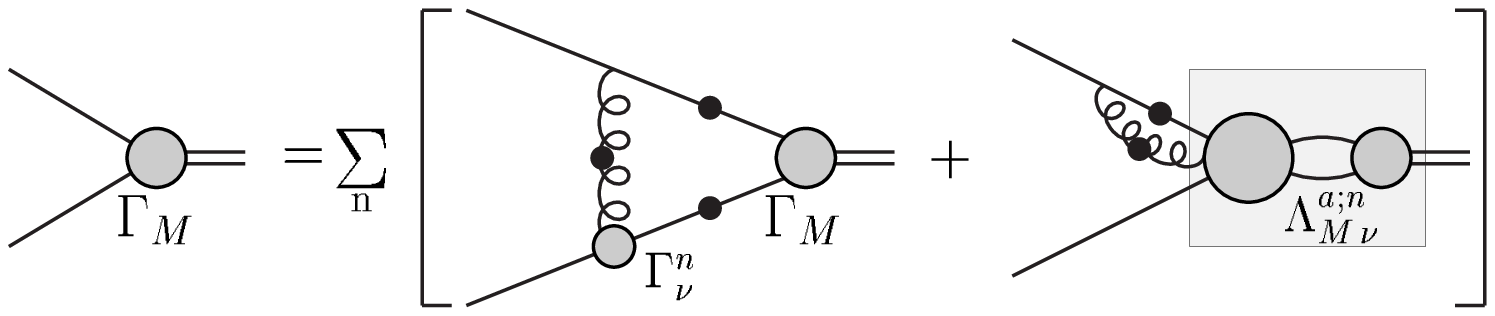}} 
\centerline{\includegraphics[width=.66\textwidth]{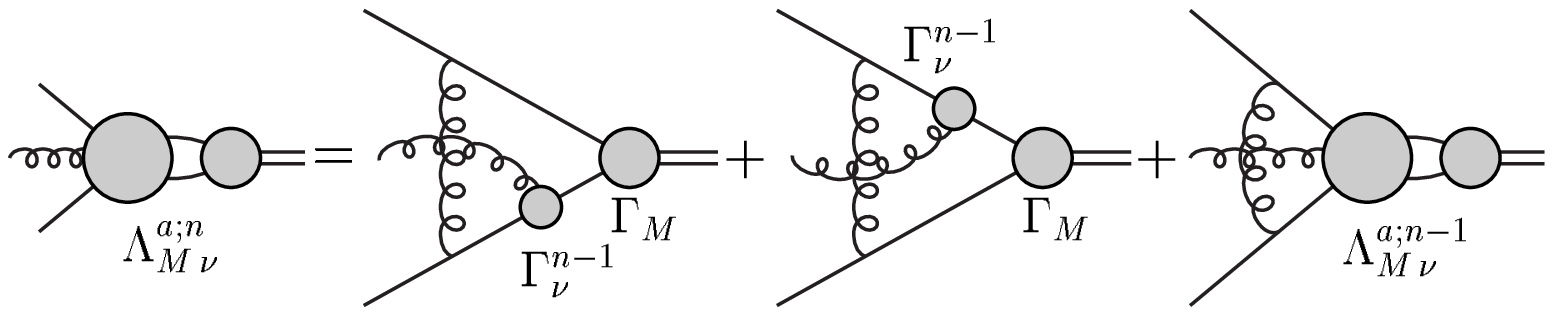}} 
\caption{\label{BSE2} Upper panel: BSE, Eq.\ (\protect\ref{genbsenL1}), which 
is valid whenever $\Gamma_\mu$ can be obtained via a recursion relation. 
Lower panel: Recursion relation for $\Lambda_{M \nu}^{a;n}$.  (Adapted from 
Ref.\ \protect\cite{detmold}.)} 
\end{figure} 
 
Equation~(\ref{genbsenL1}) is depicted in the upper panel of Fig.\ 
\ref{BSE2}.  The first term is instantly available once one has an explicit 
form for $\Gamma^{n}_{\nu}$ and the second term, identified by the shaded box 
in Fig.\ \ref{BSE2}, can be obtained \cite{detmold} via the inhomogeneous 
recursion relation depicted in the figure's lower panel.  Combining these 
figures, it is apparent that to form the Bethe-Salpeter kernel the free gluon 
line is attached to the upper dressed-quark line.  Consequently, the first 
term on the r.h.s.\ of the lower panel in Fig.\ \ref{BSE2} invariably 
generates crossed gluon lines; viz., nonplanar contributions to the kernel. 
The character of the vertex-consistent Bethe-Salpeter kernel is now clear: it 
consists of countably many contributions, a subclass of which are 
crossed-ladder diagrams and hence nonplanar.  Only the rainbow gap equation, 
obtained with $i=0$ in Eq.\ (\ref{vtxi}), yields a planar vertex-consistent 
Bethe-Salpeter kernel, namely the ladder kernel of Eq.\ (\ref{ladder}).  In 
this case alone is the number of diagrams in the dressed-vertex and kernel 
identical.  Otherwise there are always more terms in the kernel. 
 
\subsubsection{Solutions for the $\pi$- and $\rho$-mesons.}   I have 
recapitulated on a general procedure that provides the vertex-consistent 
channel-projected Bethe-Salpeter kernel once $\Gamma^{n}_{\nu}$ and the 
propagator functions; $A$, $B$, are known.  That kernel must be constructed 
independently for each channel because, e.g., $\Lambda_{M \nu}^{a}$ depends on 
$\chi_M(q;P)$.  As with the study of the vertex, an elucidation of the 
resulting BSEs' features is simplified by using the model of Eq.\ 
(\ref{mnmodel}), for then the Bethe-Salpeter kernels are finite matrices [cf.\ 
$(1 - {\cal O})^{-1}$ in Eq.\ (\ref{boldalpha})] and the homogeneous BSEs are 
merely linear, coupled algebraic equations. 
 
\begin{table}[t] 
\caption{\label{masses} Calculated $\pi$ and $\rho$ meson masses, in GeV, 
quoted with ${\cal G}= 0.48\,{\rm GeV}$, in which case $m=0.023\, {\cal G} = 
11\,$MeV.  $n$ is the number of dressed-gluon rungs retained in the planar 
vertex, see Fig.\ \protect\ref{Gamma_inf}, and hence the order of the 
vertex-consistent Bethe-Salpeter kernel: the rapid convergence of the kernel 
is apparent from the tabulated results.  (Adapted from Ref.\ 
\protect\cite{detmold}.)} 
%
\begin{center} 
\renewcommand{\arraystretch}{1.4} 
\setlength\tabcolsep{5pt} 
\begin{tabular}{l c c c c} 
\hline\noalign{\smallskip} 
 & $M_H^{n=0}$ & $M_H^{n=1}$ & $M_H^{n=2}$ & $M_H^{n=\infty}$\\[1ex]\hline 
$\pi$, $m=0$ & 0 & 0 & 0 & 0\\ 
$\pi$, $m=0.011$ & 0.152 & 0.152 & 0.152 & 0.152\\\hline 
$\rho$, $m=0$ & 0.678 & 0.745 & 0.754 & 0.754\\ 
$\rho$, $m=0.011$ & 0.695 & 0.762 & 0.770 & 0.770 \\\hline 
\end{tabular} 
\end{center} 
\end{table} 
 
Reference \cite{detmold} describes in detail the solution of the coupled gap 
and Bethe-Salpeter equations for the $\pi$- and $\rho$-mesons.  Herein I 
focus on the results, which are summarised in Table \ref{masses}.  It is 
evident that, irrespective of the order of the truncation; viz., the number 
of dressed gluon rungs in the quark-gluon vertex, the pion is massless in the 
chiral limit.  This is in spite of the fact that it is composed of heavy 
dressed-quarks, as is clear in the calculated scale of the dynamically 
generated dressed-quark mass function: see Fig.\ \ref{AM0plot}, $M(0) \approx 
{\cal G} \approx 0.5\,$GeV.  These observations emphasise that the 
masslessness of the $\pi$ is a model-independent consequence of consistency 
between the Bethe-Salpeter kernel and the kernel in the gap equation. 
Furthermore, the bulk of the $\rho$-$\pi$ mass splitting is present for $m=0$ 
and with the simplest ($n=0$; i.e., rainbow-ladder) kernel, which 
demonstrates that this mass difference is driven by the DCSB mechanism.  It 
is not the result of a carefully contrived chromo-hyperfine interaction. 
Finally, the quantitative effect of improving on the rainbow-ladder 
truncation; namely, including more dressed-gluon rungs in the gap equation's 
kernel and consistently improving the kernel in the Bethe-Salpeter equation, 
is a 10\% correction to the vector meson mass.  Simply including the first 
correction (viz., retaining the first two diagrams in Fig.\ \ref{Gamma_inf}) 
yields a vector meson mass that differs from the fully resummed result by 
$\approx 1$\%.  The rainbow-ladder truncation is clearly accurate in these 
channels. 
 
\subsubsection{Comments.} 
While I have described results obtained with a rudimentary interaction model 
in order to make the construction transparent, the procedure is completely 
general.  However, the algebraic simplicity of the analysis is naturally 
peculiar to the model.  With a more realistic interaction, the gap and vertex 
equations yield a system of twelve coupled integral equations.  The 
Bethe-Salpeter kernel for any given channel then follows as the solution of a 
determined integral equation. 
 
The material reviewed covers those points in the construction of Refs.\ 
\cite{truncscheme,detmold} that bear upon the fidelity of the rainbow-ladder 
truncation in pairing the gap equation and Bethe-Salpeter equations for the 
vector and flavour non-singlet pseudoscalar mesons.  The error is small.  In 
modelling it is therefore justified to fit one's parameters to physical 
observables at this level in these channels and then make predictions for 
other phenomena involving vector and pseudoscalar bound states in the 
expectation they will be reliable.  That approach has been successful, as 
illustrated in Ref.\ \cite{pieterrev}. 
 
Lastly, the placement of the rainbow-ladder truncation as the first term in a 
procedure that can methodically be improved explains why this truncation has 
been successful, the boundaries of its success, why it has failed outside 
these boundaries, and why sorting out the failures won't undermine the 
successes. 
 
\subsection{Selected Model-Independent Results} 
In the hadron spectrum the pion is identified as both a Goldstone mode, 
associated with DCSB, and a bound state composed of constituent $u$- and 
$d$-quarks, whose effective mass is $\sim 300 - 500\,$MeV.  Naturally, in 
quantum mechanics, one can fabricate a mass operator that yields a bound 
state whose mass is much less than the sum of the constituents' masses. 
However, that requires \textit{fine tuning} and, without additional fine 
tuning, such models predict properties for spin- and/or isospin-flip 
relatives of the pion which conflict with experiment.  A correct resolution 
of this apparent dichotomy is one of the fundamental challenges to 
establishing QCD as the theory underlying strong interaction physics, and the 
DSEs provide an ideal framework within which to achieve that end, as I now 
explain following the proof of Ref.\ \cite{mrt98}.  It cannot be emphasised 
too strongly that the legitimate understanding of pion observables; including 
its mass, decay constant and form factors, requires an approach to contain a 
well-defined and valid chiral limit. 
 
\subsubsection{Proof of Goldstone's Theorem.} 
Consider the BSE expressed for the isovector pseudoscalar channel: 
\begin{equation} 
\label{genbsepi} 
\left[\Gamma_\pi^j(k;P)\right]_{tu} =  \int^\Lambda_q 
\,[\chi_\pi^j(q;P)]_{sr} \,K^{rs}_{tu}(q,k;P)\,, 
\end{equation} 
with $\chi_\pi^j(q;P)=S(q_+) \Gamma_\pi^j(q;P) S(q_-)$ obvious from Eq.\ 
(\ref{chiM}) and $j$ labelling isospin, of which the solution has the general 
form 
\begin{eqnarray} 
\nonumber 
\Gamma_\pi^j(k;P) & = &  \tau^j \gamma_5 \left[ i E_\pi(k;P) + 
\gamma\cdot P F_\pi(k;P) \rule{0mm}{5mm}\right. \\ 
& & \left. \rule{0mm}{5mm}+ \gamma\cdot k \,k \cdot P\, G_\pi(k;P) + 
\sigma_{\mu\nu}\,k_\mu P_\nu \,H_\pi(k;P) \right]. \label{genpibsa} 
\end{eqnarray} 
It is apparent that the dressed-quark propagator, the solution of Eq.\ 
(\ref{gendse}), is an important part of the BSE's kernel. 
 
Chiral symmetry and its dynamical breaking are expressed in the axial-vector 
Ward-Takahashi identity, Eq.\ (\ref{avwti}), which involves the axial-vector 
vertex: 
\begin{equation} 
\label{genave} 
\left[\Gamma_{5\mu}^j(k;P)\right]_{tu} = 
Z_2 \, \left[\gamma_5\gamma_\mu \frac{\tau^j}{2}\right]_{tu} \,+ 
\int^\Lambda_q \, [\chi_{5\mu}^j(q;P)]_{sr} \,K^{rs}_{tu}(q,k;P)\,, 
\end{equation} 
that has the 
general form 
\begin{eqnarray} 
\nonumber\Gamma_{5 \mu}^j(k;P) & = & 
\frac{\tau^j}{2} \gamma_5 
\left[ \gamma_\mu F_R(k;P) + \gamma\cdot k k_\mu G_R(k;P) 
- \sigma_{\mu\nu} \,k_\nu\, H_R(k;P) 
\right]\\ 
& + & 
 \tilde\Gamma_{5\mu}^{j}(k;P) 
+ \frac{P_\mu}{P^2 + m_\phi^2} \phi^j(k;P)\,, 
\label{genavv} 
\end{eqnarray} 
where $F_R$, $G_R$, $H_R$ and $\tilde\Gamma_{5\mu}^{i}$ are regular as 
$P^2\to -m_\phi^2$, $P_\mu \tilde\Gamma_{5\mu}^{i}(k;P) \sim {\rm O }(P^2)$ 
and $\phi^j(k;P)$ has the structure depicted in Eq.\ (\ref{genpibsa}). 
Equation (\ref{genavv}) admits the possibility of at least one pole term in 
the vertex but does not require it. 
 
Substituting Eq.\ (\ref{genavv}) into (\ref{genave}) and equating putative 
pole terms, it is clear that, if present, $\phi^j(k;P)$ satisfies Eq.\ 
(\ref{genbsepi}).  Since this is an eigenvalue problem that only admits a 
$\Gamma_\pi^j \neq 0$ solution for $P^2= -m_\pi^2$, it follows that 
$\phi^j(k;P)$ is nonzero solely for $P^2= -m_\pi^2$ and the pole mass is 
$m_\phi^2 = m_\pi^2$.  Hence, if $K$ supports such a bound state, the 
axial-vector vertex contains a pion-pole contribution.  Its residue, $r_A$, 
however, is not fixed by these arguments.  Thus Eq.\ (\ref{genavv}) becomes 
\begin{eqnarray} 
\nonumber 
\Gamma_{5 \mu}^j(k;P) & = & 
\frac{\tau^j}{2} \gamma_5 
\left[ \gamma_\mu F_R(k;P) + \gamma\cdot k k_\mu G_R(k;P) 
- \sigma_{\mu\nu} \,k_\nu\, H_R(k;P) \right]\\ 
& & + 
 \tilde\Gamma_{5\mu}^{i}(k;P) 
+ \frac{r_A P_\mu}{P^2 + m_\pi^2} \Gamma_\pi^j(k;P)\,. 
\label{truavv} 
\end{eqnarray} 
 
Consider now the chiral limit axial-vector Ward-Takahashi identity, Eq.\ 
(\ref{avwti}).  If one assumes $m_\pi^2=0$ in Eq.\ (\ref{truavv}), 
substitutes it into the l.h.s.\ of Eq.\ (\ref{avwti}) along with Eq.\ 
(\ref{sinvp}) on the right, and equates terms of order $(P_\nu)^0$ and 
$P_\nu$, one obtains the chiral-limit relations \cite{mrt98} 
\begin{equation} 
\label{bfgwti} 
\begin{array}{rclcrcl} 
r_A E_\pi(k;0)  &= &  B(k^2)\,, 
&\;\; & F_R(k;0) +  2 \, r_A F_\pi(k;0) & = & A(k^2)\,, \\ 
G_R(k;0) +  2 \,r_A G_\pi(k;0)    & = & 2 A^\prime(k^2)\,, 
&\;\; & H_R(k;0) +  2 \,r_A H_\pi(k;0)    & = & 0\,. 
\end{array} 
\end{equation} 
I have already explained that $B(k^2) \equiv 0$ in the chiral limit [remember 
Eq.\ (\ref{Bpert0})] and that a $B(k^2)\neq 0$ solution of Eq.\ 
(\ref{gendse}) in the chiral limit signals DCSB.  Indeed, in this case 
\cite{LanePolitzer} 
\begin{equation} 
\label{Mchiral} 
M(p^2) \stackrel{{\rm large}-p^2}{=}\, 
\frac{2\pi^2\gamma_m}{3}\,\frac{\left(-\,\langle \bar q q \rangle^0\right)} 
           {p^2 
        \left(\sfrac{1}{2}\ln\left[p^2/\Lambda_{\rm QCD}^2\right] 
        \right)^{1-\gamma_m}}\,, 
\end{equation} 
where $\langle \bar q q \rangle^0$ is the renormalisation-point-independent 
vacuum quark condensate \cite{bankscasher}.  Furthermore, there is at least 
one nonperturbative DSE truncation scheme that preserves the axial-vector 
Ward-Takahashi identity, order by order.  Hence Eqs. (\ref{bfgwti}) are exact 
quark-level Goldberger-Treiman relations, which state that when chiral 
symmetry is dynamically broken:\vspace*{-\baselineskip} 
\begin{center} 
\parbox{32em}{\flushleft 
\renewcommand{\theenumi}{(\roman{enumi})} 
\begin{enumerate} 
\item the homogeneous isovector pseudoscalar BSE has a massless solution; 
%
\item the Bethe-Salpeter amplitude for the massless bound state has a term 
proportional to $\gamma_5$ alone, with $E_\pi(k;0)$ completely determined by 
the scalar part of the quark self energy, in addition to other pseudoscalar 
Dirac structures, $F_\pi$, $G_\pi$ and $H_\pi$, that are nonzero; 
%
\item and the axial-vector vertex is dominated by the pion pole for 
$P^2\simeq 0$. 
\end{enumerate}} 
\end{center} 
The converse is also true.  Hence DCSB is a sufficient and necessary 
condition for the appearance of a massless pseudoscalar bound state (of what 
can be very-massive constituents) that dominates the axial-vector vertex for 
$P^2\approx 0$. 
 
\subsubsection{Mass Formula.} 
\label{sec:massformula} 
When chiral symmetry is explicitly broken the axial-vector Ward-Takahashi 
identity becomes: 
\begin{equation} 
\label{avwtim} 
P_\mu \Gamma_{5\mu}^j(k;P)  = S^{-1}(k_+) i \gamma_5\frac{\tau^j}{2} 
+  i \gamma_5\frac{\tau^j}{2} S^{-1}(k_-) 
- 2i\,m(\zeta) \,\Gamma_5^j(k;P) , 
\end{equation} 
where the pseudoscalar vertex is obtained from 
\begin{eqnarray} 
\label{genpve} 
\left[\Gamma_{5}^j(k;P)\right]_{tu} & = & 
Z_4\,\left[\gamma_5 \frac{\tau^j}{2}\right]_{tu} \,+ 
\int^\Lambda_q \, 
\left[ \chi_5^j(q;P)\right]_{sr} 
K^{rs}_{tu}(q,k;P)\,. 
\end{eqnarray} 
As argued in connection with Eq.\ (\ref{genave}), the solution of Eq.\ 
(\ref{genpve}) has the form 
\begin{eqnarray} 
 \nonumber 
i \Gamma_{5 }^j(k;P) & = & 
\frac{\tau^j}{2} \gamma_5 
\left[ i E_R^P(k;P) + \gamma\cdot P \, F_R^P 
+ \gamma\cdot k \,k\cdot P\, G_R^P(k;P) 
\right. \\ 
& & 
\left.+ \, \sigma_{\mu\nu}\,k_\mu P_\nu \,H_R^P(k;P) \right] 
+ \frac{ r_P }{P^2 + m_\pi^2} \Gamma_\pi^j(k;P)\,,\label{genpvv} 
\end{eqnarray} 
where $E_R^P$, $F_R^P$, $G_R^P$ and $H_R^P$ are regular as $P^2\to -m_\pi^2$; 
i.e., the isovector pseudoscalar vertex also receives a contribution from the 
pion pole.  In this case equating pole terms in the Ward-Takahashi identity, 
Eq.\ (\ref{avwtim}), entails \cite{mrt98} 
\begin{equation} 
\label{gmora} 
r_A \,m_\pi^2 = 2\,m(\zeta) \,r_P(\zeta)\,. 
\end{equation} 
This, too, is an exact relation in QCD.  Now it is important to determine the 
residues $r_A$ and $r_P$. 
 
Study of the renormalised axial-vector vacuum polarisation shows 
\cite{mrt98}: 
\begin{eqnarray} 
\label{fpiexact} r_A \,\delta^{ij} \,  P_\mu = f_\pi \,\delta^{ij} \,  P_\mu 
&=& Z_2\,{\rm tr} \int^\Lambda_q \sfrac{1}{2} \tau^i \gamma_5\gamma_\mu S(q_+) 
\Gamma_\pi^j(q;P) S(q_-)\,, 
\end{eqnarray} 
where the trace is over colour, Dirac and flavour indices; i.e., the residue 
of the pion pole in the axial-vector vertex is the pion decay constant.  The 
factor of $Z_2$ on the r.h.s.\ in Eq.\ (\ref{fpiexact}) is crucial: it 
ensures the result is gauge invariant, and cutoff and renormalisation-point 
independent.  Equation (\ref{fpiexact}) is the exact expression in quantum 
field theory for the pseudovector projection of the pion's wave function on 
the origin in configuration space. 
 
A close inspection of Eq.\ (\ref{genpve}), following its re-expression in 
terms of the renormalised, fully-amputated quark-antiquark scattering 
amplitude: $M = K + K (SS) K + \ldots$, yields \cite{mrt98} 
\begin{equation} 
\label{cpres} i r_P\, \delta^{ij}  = Z_4\,{\rm tr} \int^\Lambda_q \sfrac{1}{2} 
\tau^i \gamma_5 S(q_+) \Gamma_\pi^j(q;P) S(q_-)\,, 
\end{equation} 
wherein the dependence of $Z_4$ on the gauge parameter, the regularisation 
mass-scale and the renormalisation point is exactly that required to ensure: 
1) $r_P$ is finite in the limit $\Lambda\to \infty$; 2) $r_P$ is 
gauge-parameter independent; and 3) the renormalisation point dependence of 
$r_P$ is just such as to guarantee the r.h.s.\ of Eq.\ (\ref{gmora}) is 
renormalisation point \textit{independent}.  Equation (\ref{cpres}) expresses 
the pseudoscalar projection of the pion's wave function on the origin in 
configuration space. 
 
Focus for a moment on the chiral limit behaviour of Eq.\ (\ref{cpres})
whereat, using Eqs.\ (\ref{genpibsa}), (\ref{bfgwti}), one finds readily
\begin{equation} 
- \langle \bar q q \rangle_\zeta^0 = f_\pi r_P^0(\zeta) 
= Z_4(\zeta,\Lambda)\, N_c\, {\rm tr}_{\rm D} \int^\Lambda_q \! S_{\hat m 
=0}(q) \,. \label{qbq0} 
\end{equation} 
Equation (\ref{qbq0}) is unique as the expression for the chiral limit 
\textit{vacuum quark condensate}.\footnote{The trace of the massive 
dressed-quark propagator is not renormalisable and hence there is no unique 
definition of a massive-quark condensate \protect\cite{bankscasher}.}  It is 
$\zeta$-dependent but independent of the gauge parameter and the 
regularisation mass-scale, and Eq.\ (\ref{qbq0}) thus proves that the 
chiral-limit residue of the pion pole in the pseudoscalar vertex is$\,$ $ 
(-\langle \bar q q \rangle_\zeta^0) /f_\pi$.  Now Eqs.\ (\ref{gmora}), 
(\ref{qbq0}) yield 
\begin{equation} 
\label{gmor} 
(f_\pi^0)^2 \, m_\pi^2 = - 2 \, m(\zeta)\, \langle \bar q q \rangle_\zeta^0 + 
{\rm O}(\hat m^2)\,, 
\end{equation} 
where $f_\pi^0$ is the chiral limit value from Eq.\ (\ref{fpiexact}).  Hence 
what is commonly known as the Gell-Mann--Oakes--Renner relation is a 
\textit{corollary} of Eq.\ (\ref{gmora}). 
 
One can now understand the results in Table \ref{masses}: a massless bound 
state of massive constituents is a necessary consequence of DCSB and will 
emerge in any few-body approach to QCD that employs a systematic truncation 
scheme which preserves the Ward-Takahashi identities. 
 
Upon review it will be apparent that Eqs.\ (\ref{gmora}) -- (\ref{cpres}) are 
valid for any values of the current-quark masses, and the generalisation to 
$N_f$ quark flavours is \cite{mr97,marisAdelaide,mishaSVY} 
\begin{equation} 
\label{massformula} 
f_H^2 \, m_H^2 = - \, \langle \bar q q \rangle_\zeta^H {\cal M}_H^\zeta, 
\end{equation} 
${\cal M}^H_\zeta = m^\zeta_{q_1} + m^\zeta_{q_2}$ is the sum of the 
current-quark masses of the meson's constituents; 
\begin{equation} 
\label{fH} 
f_H \, P_\mu = Z_2 \, {\rm tr} \int_q^\Lambda \! \sfrac{1}{2} (T^H)^T 
\gamma_5 \gamma_\mu {\cal S}(q_+)\, \Gamma^H(q;P)\, {\cal S}(q_-)\,, 
\end{equation} 
with ${\cal S}= {\rm diag}(S_u,S_d,S_s,\ldots)$, $T^H$ a flavour matrix 
specifying the meson's quark content, e.g., $T^{\pi^+}=\sfrac{1}{2} 
(\lambda^1+i\lambda^2)$, $\{\lambda^i\}$ are $N_f$-flavour generalisations of 
the Gell-Mann matrices; and 
\begin{equation} 
\label{qbqH} 
\langle \bar q q \rangle^H_\zeta = - f_H\,r_H^\zeta = i f_H\, Z_4\, {\rm 
tr}\int_q^\Lambda \!  \sfrac{1}{2}(T^H)^T \gamma_5 {\cal S}(q_+) 
\,\Gamma^H(q;P)\, {\cal S}(q_-) \,. 
\end{equation} 
NB.\ Equation (\ref{qbq0}) means that in the chiral limit $\langle \bar q q 
\rangle^H_\zeta \to \langle \bar q q \rangle_\zeta^0$ and hence $\langle \bar 
q q \rangle^H_\zeta$ has been called an \textit{in-hadron condensate}. 
 
The formulae reviewed in this Section also yield model-independent 
corollaries for systems involving heavy-quarks, as I relate in Sec.\ 
\ref{sec:HQ}. 
 
\section{Basis for a Description of Mesons} 
\label{sect4label} 
The \label{sect3label} renormalisation-group-improved rainbow-ladder 
truncation has long been employed to study light mesons, and in Secs.\ \ref 
{subsec:truncation}, \ref{subsec:systematic} it was shown to be a 
quantitatively reliable tool for vector and flavour nonsinglet pseudoscalar 
mesons.  In connection with Eqs.\ (\ref{ladder}), (\ref{rainbowdse}), I 
argued that the truncation preserves the ultraviolet behaviour of the 
quark-antiquark scattering kernel in QCD but requires an assumption about 
that kernel in the infrared; viz., on the domain $Q^2 \lsim 1\,$GeV$^2$, 
which corresponds to length-scales $\gsim 0.2\,$fm.  The calculation of this 
behaviour is a primary challenge in contemporary hadron physics and there is 
progress 
\cite{raya,cdrvienna,latticegluon,blochmrgluon,fischer,latticevertex,langfeld}. 
However, at present the most efficacious approach is to model the kernel in 
the infrared, which enables quantitative comparisons with experiments that 
can be used to inform theoretical analyses.  The most extensively applied 
model is specified by \cite{pmspectra2} 
\begin{equation} 
\frac{\alpha(Q^2)}{Q^2} = \frac{4\pi^2}{\omega^6} D\, Q^2 {\rm 
e}^{-Q^2/\omega^2} + \, \frac{ 8\pi^2\, \gamma_m } { \ln\left[\tau + \left(1 + 
Q^2/\Lambda_{\rm QCD}^2\right)^2\right]} \, {\cal F}(Q^2) \,, \label{gk2} 
\end{equation} 
in Eqs.\ (\ref{ladder}), (\ref{rainbowdse}).  Here, ${\cal F}(Q^2)= [1 -
\exp(-Q^2/[4 m_t^2])]/Q^2$, $m_t$ $=$ $0.5\,$GeV; $\tau={\rm e}^2-1$;
$\gamma_m = 12/25$; and \cite{pdg98} $\Lambda_{\rm QCD} =
\Lambda^{(4)}_{\overline{\rm MS}}=0.234\,$GeV.  This simple form expresses
the interaction strength as a sum of two terms: the second ensures that
perturbative behaviour is preserved at short-range; and the first makes
provision for the possibility of enhancement at long-range.  The true
parameters in Eq.\ (\ref{gk2}) are $D$ and $\omega$, which together determine
the integrated infrared strength of the rainbow-ladder kernel; i.e., the
so-called interaction tension, $\sigma^\Delta$ \cite{cdrvienna}. However, I
emphasise that they are not independent: in fitting to a selection of
observables, a change in one is compensated by altering the other; e.g., on
the domain $\omega\in[0.3,0.5]\,$GeV, the fitted observables are
approximately constant along the trajectory \cite{raya}
\begin{equation} 
\label{omegaD} 
\omega \,D = (0.72\,{\rm GeV})^3. 
\end{equation} 
Hence Eq.\ (\ref{gk2}) is a one-parameter model.  This correlation: a 
reduction in $D$ compensating an increase in $\omega$, ensures a fixed value 
of the interaction tension. 
 
\subsection{Rainbow Gap Equation} 
Equations (\ref{rainbowdse}) and (\ref{gk2}) provide a model for QCD's gap
equation and in hadron physics applications one is naturally interested in
the nonperturbative DCSB solution.  A familiar property of gap equations is
that they only support such a solution if the interaction tension exceeds
some critical value.  In the present case that value is $\sigma_c^\Delta \sim
2.5\,$GeV/fm \cite{cdrvienna}.  This amount of infrared strength is
sufficient to generate a nonzero vacuum quark condensate \textit{but only
just}.  An acceptable description of hadrons requires $\sigma^\Delta \sim
25\,$GeV/fm \cite{mr97} and that is obtained with \cite{pmspectra2}
\begin{equation} 
\label{valD} 
D= (0.96\,{\rm GeV})^2 \,. 
\end{equation} 
 
This value of the model's infrared mass-scale parameter and the two 
current-quark masses 
\begin{equation} 
\label{mqs} 
\begin{array}{cc} 
m_u(1\,{\rm GeV})=5.5\, {\rm MeV} \,,\;&  m_s(1\,{\rm GeV})=125\, {\rm MeV}\,, 
\end{array} 
\end{equation} 
defined using the one-loop expression 
\begin{equation} 
\label{Zmone} 
\frac{m(\zeta)}{m(\zeta^\prime)} = Z_m(\zeta^\prime,\zeta) \stackrel{\rm 
1-loop}{=} \left( \frac{\ln[\zeta^\prime/\Lambda_{\rm 
QCD}]}{\ln[\zeta/\Lambda_{\rm QCD}]}\right)^{\gamma_m} 
\end{equation} 
to evolve $m_u(19\,{\rm GeV})=3.7\,$MeV and $m_s(19\,{\rm GeV})=85\,$MeV,
were obtained in Ref.\ \cite{pmspectra2} by requiring a least-squares fit to
the $\pi$- and $K$-meson observables listed in Table \ref{RGIpiK}.  The
procedure was straightforward: the rainbow gap equation [Eqs.\
(\ref{renormS}), (\ref{rainbowdse}), (\ref{gk2})] was solved with a given
parameter set and the output used to complete the kernels in the homogeneous
ladder BSEs for the $\pi$- and $K$-mesons [Eqs.\ (\ref{ladder}),
(\ref{rainbowdse}), (\ref{genbsepi}), (\ref{genpibsa}) with $\tau^j$ for the
$\pi$ channel and $\tau^j \to T^{K^+} = \sfrac{1}{2}(\lambda^4+i\lambda^5)$
for the $K$].  These BSEs were solved to obtain the $\pi$- and $K$-meson
masses, and the Bethe-Salpeter amplitudes.  Combining this information
delivers the leptonic decay constants via Eq.\ (\ref{fH}).  This was repeated
as necessary to arrive at the results in Table \ref{RGIpiK}.  The model gives
a vacuum quark condensate
\begin{equation} 
\label{qbq0val} 
-\langle \bar q q \rangle^0_{1\,{\rm GeV}}= (0.242\,{\rm GeV})^3\,, 
\end{equation} 
calculated from Eq.\ (\protect\ref{qbq0}) and evolved using the one-loop 
expression in Eq.\ (\ref{Zmone}). 
 
\begin{table}[t] 
\caption{\label{RGIpiK} Comparison of experimental values with results for 
$\pi$ and $K$ observables calculated using the rainbow-ladder interaction 
specified by Eq.\ (\protect\ref{gk2}), quoted in MeV.  The model's sole 
parameter and the current-quark masses were varied to obtain these results, 
and the best fit parameter values are given in Eqs.\ (\protect\ref{valD}), 
(\protect\ref{mqs}).  Predictions for analogous vector meson observables are 
also tabulated.  NB.\ A charged particle normalisation is used for $f_H^V$ in 
Eq.\ (\protect\ref{fV}), which differs from that in Eq.\ (\protect\ref{fH}) 
by a multiplicative factor of $\surd 2$.  (Adapted from Ref.\ 
\protect\cite{pmspectra2}.)} 
\begin{center} 
\renewcommand{\arraystretch}{1.4} 
\setlength\tabcolsep{5pt} 
\begin{tabular}{l | c c c c||c c c c c c } 
\hline\noalign{\smallskip} 
         & $m_\pi$ & $m_K$ & $f_\pi$ & $f_K$ 
& $m_\rho$ & $m_{K^\ast}$ & $m_{\phi}$ 
         & $f_\rho$ & $f_{K^\ast}$ & $f_\phi$  \\\hline 
Calc.\ \protect\cite{pmspectra2} & 138 & 497 & 93 & 109 \rule{0ex}{2.7ex} 
& 742 & 936 & 1072 & 207 & 241 & 259 \\ 
Expt.\ \protect\cite{pdg98} & 138 & 496 & 92 & 113 
& 771 & 892 & 1019 & 217 & 227 & 228 \\\hline 
Rel.\ Error   &&&& & 0.04 & -0.05 & -0.05 & 0.05 & -0.06 & -0.14 \\\hline 
 
\end{tabular} 
\end{center} 
\end{table} 
 
With the model's single parameter fixed, and the dressed-quark propagator 
determined, it is straightforward to compose and solve the homogeneous BSE 
for vector mesons.  This yields predictions, also listed in Table 
\ref{RGIpiK}, for the vector meson masses and electroweak decay constants 
\cite{mishaSVY} 
\begin{equation} 
\label{fV} f_H^V \, M_H^V = \sfrac{1}{3} Z_2 \, {\rm tr} \int_q^\Lambda \! 
(T^H)^T \gamma_\mu\, {\cal S}(q_+)\, \Gamma_\mu^H(q;P)\, {\cal S}(q_-)\,, 
\end{equation} 
where $M_H^V$ is the meson's mass and $P_\mu \Gamma_\mu^H(q;P)=0$ for 
$P^2=-(M_V^H)^2$; i.e., the Bethe-Salpeter amplitude is transverse.  $f_H^V$ 
characterises decays such as $\rho\to e^+ e^-$, $\tau \to K^\ast \nu_\tau$. 
 
Given the discussion in Sec.\ \ref{sect2label}, the phenomenological success 
of the rainbow-ladder kernel, manifest in the results of Table \ref{RGIpiK}, 
is unsurprising and, indeed, was to be expected. 
 
\subsection{Comparison with Lattice Simulations} 
\begin{figure}[t] 
\centerline{\includegraphics[width=.51\textwidth]{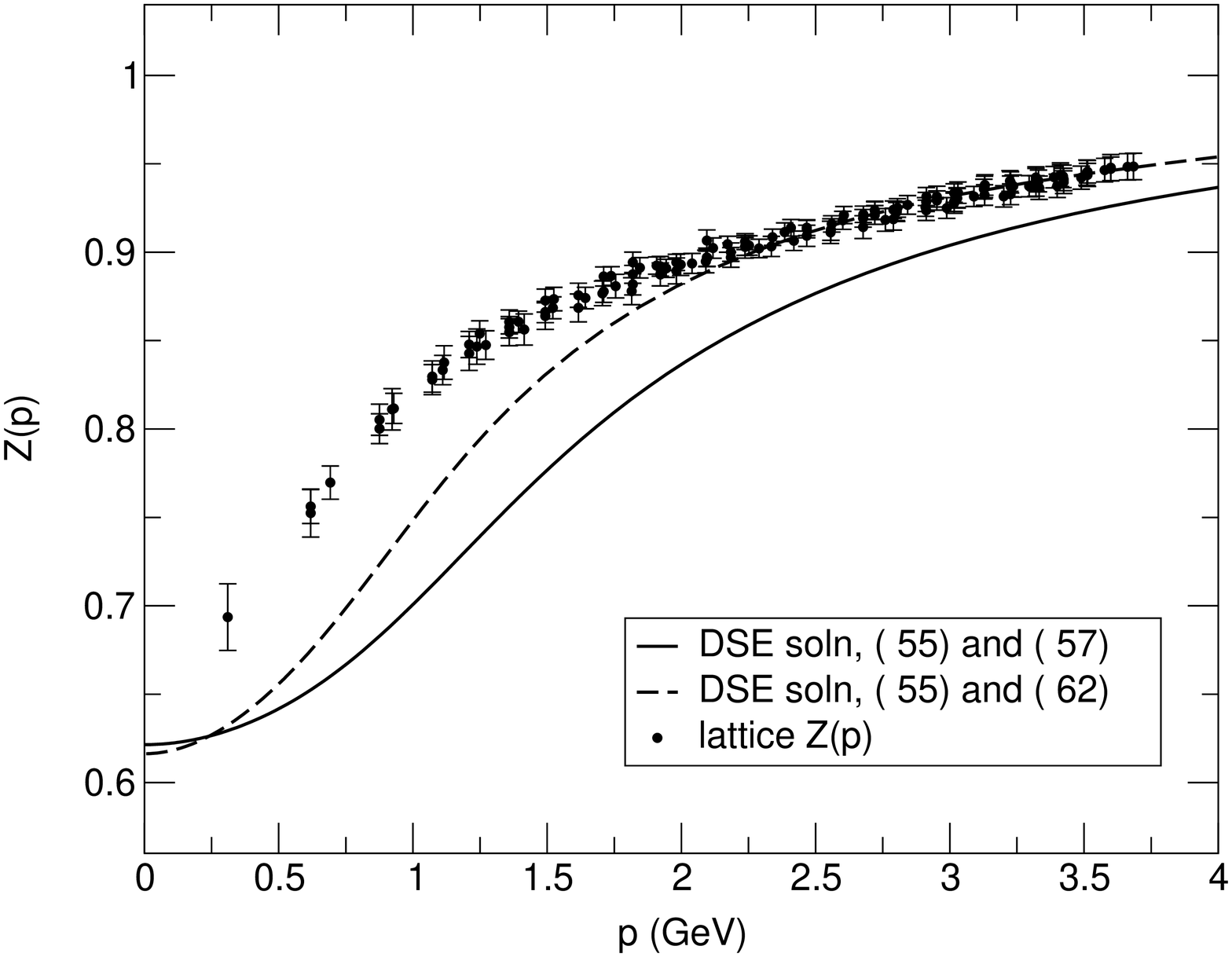} 
\includegraphics[width=.51\textwidth]{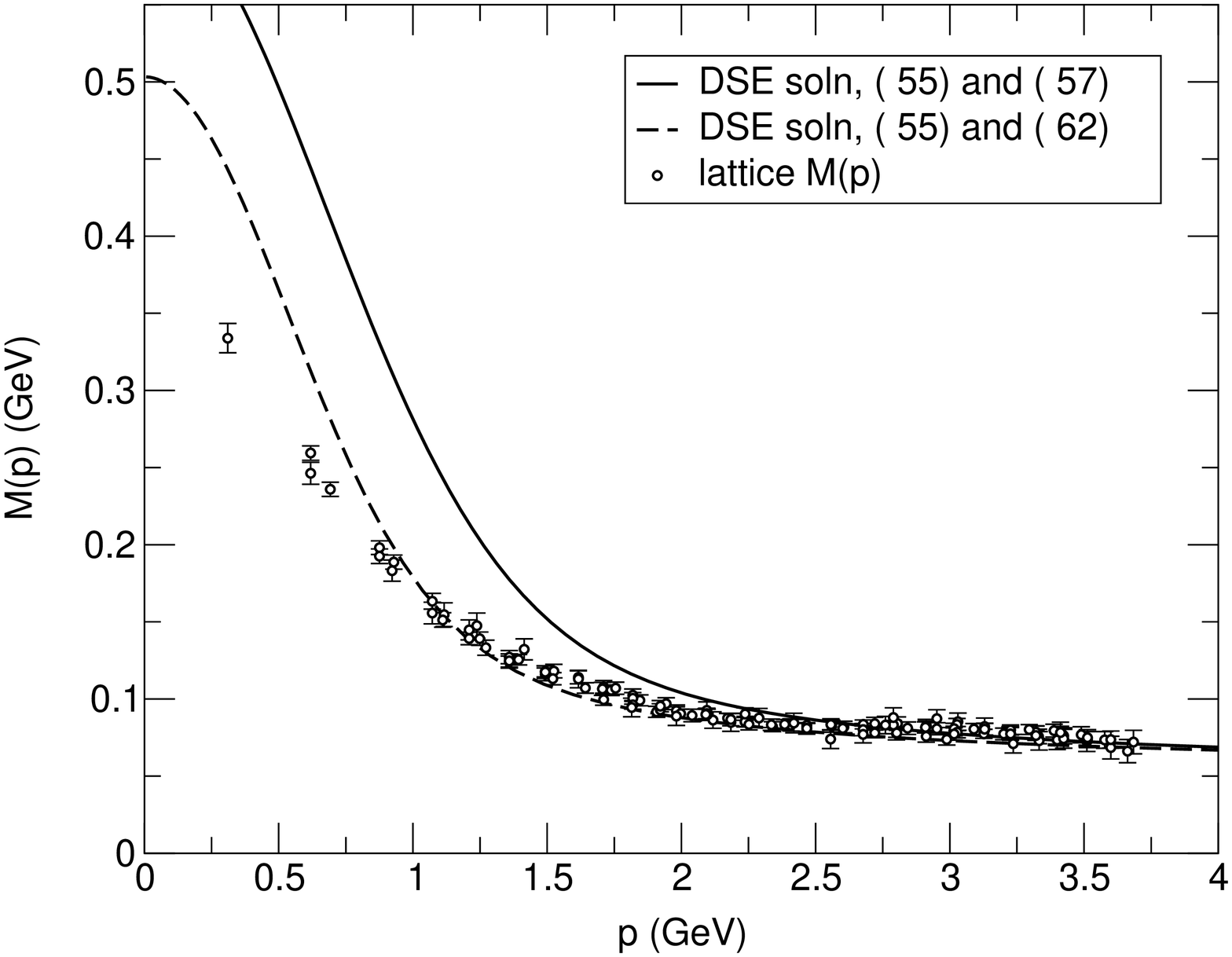}} 
\caption{\label{figZ} Left panel -- wave function renormalisation: solution 
of the gap equation using Eqs.\ (\protect\ref{gk2}), (\protect\ref{valD}) 
(\textit{solid curve}); solution using Eqs.\ (\protect\ref{gk2}), 
(\protect\ref{DomegaL}) (\textit{dashed-curve}); quenched lattice-QCD 
simulations \protect\cite{latticequark}, obtained with $m=0.036/a\sim 
60\,$MeV (\textit{data}). The DSE study used a renormalisation point 
$\zeta=19\,$GeV and a current-quark mass $0.6\,m_s^{1\,{\rm GeV}}$ [Eq.\ 
(\protect\ref{mqs})], to enable a direct comparison with the lattice data. 
Right panel -- mass function. 
(Adapted from Ref.\ \protect\cite{raya}.)} 
\end{figure}

The solution of the gap equation has long been of interest in grappling with 
DCSB in QCD and hence, in Figs.\ \ref{figZ}, I depict the scalar functions 
characterising the renormalised dres\-sed-quark propagator: the wave function 
renormalisation, $Z(p^2)$, and mass function, $M(p^2)$, obtained by solving 
Eq.\ (\ref{rainbowdse}) using Eq.\ (\ref{gk2}).  The infrared suppression of 
$Z(p^2)$ and enhancement of $M(p^2)$ are longstanding predictions of DSE 
studies \cite{cdragw}.  Indeed, this property of asymptotically free theories 
was elucidated in Refs.\ \cite{LanePolitzer} and could be anticipated from 
studies of strong coupling QED \cite{bjw}.  The prediction has recently been 
confirmed in numerical simulations of quenched lattice-QCD, as is evident in 
the figures. 
 
It is not yet possible to reliably determine the behaviour of lattice 
Schwinger functions for current-quark masses that are a realistic 
approximation to those of the $u$- and $d$-quarks.  A veracious lattice 
estimate of $m_\pi$, $f_\pi$, $\langle \bar q q\rangle^0$ is therefore 
absent.  To obtain such an estimate, Ref.\ \cite{raya} used the rainbow 
kernel described herein and varied $(D,\omega)$ in order to reproduce the 
quenched lattice-QCD data.  A best fit was obtained with 
\begin{equation} 
\label{DomegaL} 
\begin{array}{cc} 
D=(0.74\,{\rm GeV})^2\,,\; & \omega=0.3\,{\rm GeV}\,, 
\end{array} 
\end{equation} 
at a current-quark mass of $0.6\,m_s^{1\,{\rm GeV}}\!\approx 14\,m_u$ [Eq.\ 
(\protect\ref{mqs})] chosen to coincide with that employed in the lattice 
simulation.  Constructing and solving the homogeneous BSE for a pion-like bound 
state composed of quarks with this current-mass yields 
\begin{equation} 
m_{\pi}^{m_q \sim 14 m_u} = 0.48\,{\rm GeV},\; f_{\pi}^{m_q \sim 14 m_u} = 
0.094\,{\rm GeV}\,. 
\end{equation} 
The parameters in Eq.~(\ref{DomegaL}) give chiral limit results \cite{raya}: 
\begin{equation} 
\label{latticechiral} 
f_\pi^0= 0.068\,{\rm GeV}\,,\; 
-\langle\bar q q\rangle^0_{1 \,{\rm GeV}} = (0.19\,{\rm GeV})^3\,, 
\end{equation} 
whereas Eqs.\ (\ref{omegaD}), (\ref{valD}) give $f_\pi^0=0.088\,$GeV.  These 
results have been confirmed in a more detailed analysis \cite{mandarlattice} 
and this correspondence suggests that chiral and physical pion observables 
are materially underestimated in the quenched theory: $|\langle \bar q q 
\rangle|$ by a factor of two and $f_\pi$ by $30$\%. 
 
The rainbow-ladder kernel has also been employed in an analysis of a 
trajectory of fictitious pseudoscalar mesons, all composed of equally massive 
constituents \cite{pmqciv} (The only physical state on this trajectory is the 
pion.)  The DSE study predicts \cite{cdrqciv} 
\begin{equation} 
\frac{m_{H_{m=2 m_s}}}{m_{H_{m=m_s}}} = 2.2\,, 
\end{equation} 
in agreement with a result of recent quenched lattice simulations
\cite{michaels}.  The DSE study provides an intuitive understanding of this
result, showing that it owes itself to a large value of the in-hadron
condensates for light-quark mesons; e.g., $\langle \bar q q\rangle^{s\bar
s}_{1\,{\rm GeV}}= (-0.32\,{\rm GeV})^3$ \cite{mr97}, and thereby confirms
the large-magnitude condensate version of chiral perturbation theory, an
observation also supported by Eq.\ (\ref{latticechiral}). References
\cite{pmqciv,TandyErice} also provide vector meson trajectories.
 
\subsection{{\it Ab Initio} Calculation of Meson Properties} 
\label{abinitio} 
The renormalisation-group-improved rainbow-ladder kernel defined with Eq.\ 
(\ref{gk2}) has been employed to predict a wide range of meson observables, 
and this is reviewed in Ref.\ \cite{pieterrev}.  These results; e.g., those 
for vector mesons in Table \ref{RGIpiK}, are true predictions, in the sense 
that the model's mass-scale was fixed, as described in connection with Eq.\ 
(\ref{valD}), and every element in each calculation was completely determined 
by, and calculated from, that kernel. 
 
\begin{figure}[t] 
\centerline{\includegraphics[width=.60\textwidth]{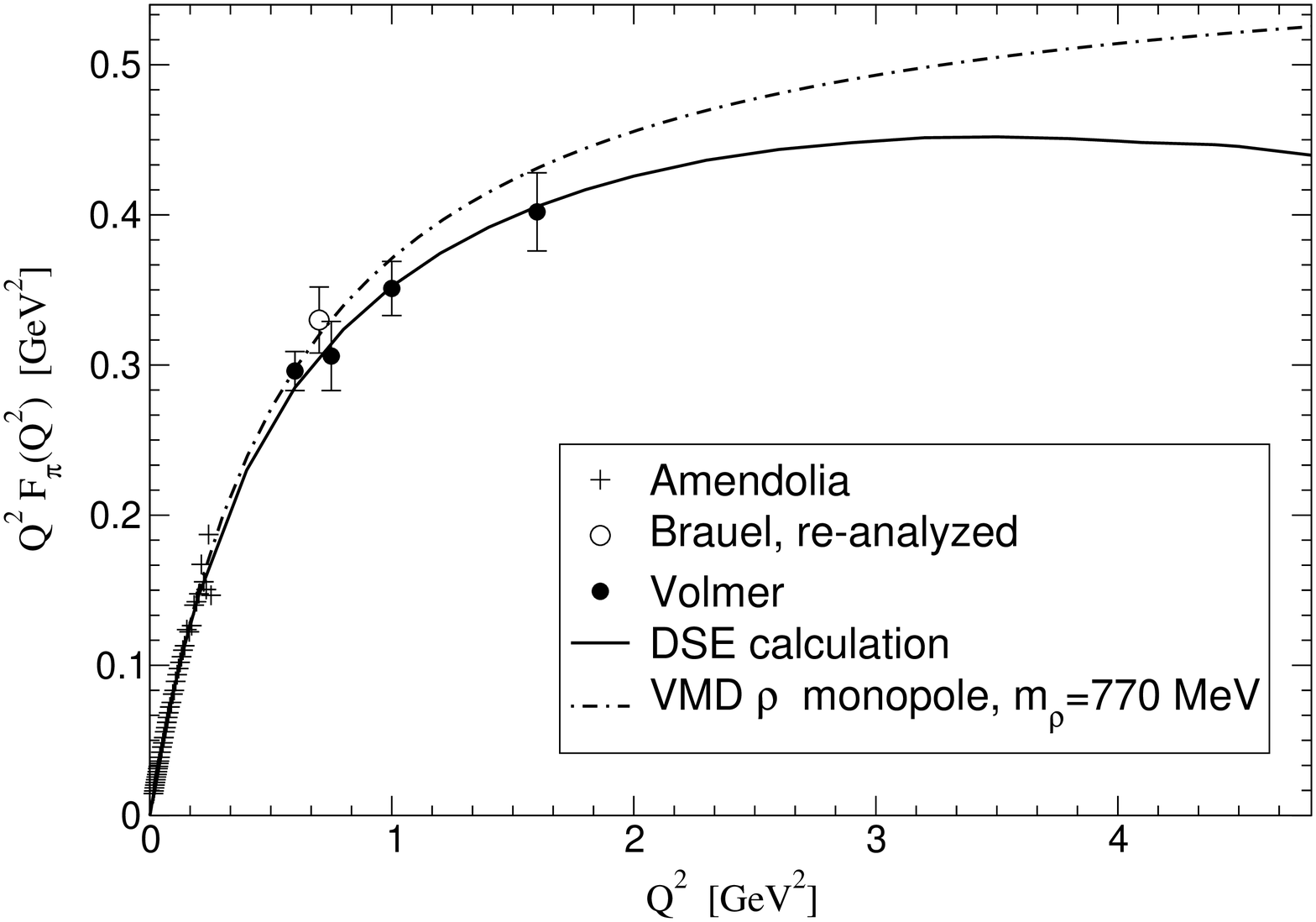}} 
\caption{\label{fig:fpi} Impulse approximation DSE prediction for $q^2 
F_\pi(q^2)$ obtained in a parameter-free application of the 
renormalisation-group-improved rainbow-ladder truncation, Eqs.\ 
(\protect\ref{gk2}), (\protect\ref{valD}). The data are from Refs.\ 
\protect\cite{Brauel:1979zk,Amendolia:1986wj,Volmer:2000ek}. (Adapted from 
Ref.\ \protect\cite{Maris:2001am}.)} 
\end{figure}

A particular success was the calculation of the electromagnetic pion form 
factor, which is described in Refs.\ \cite{pieterpion,pieterpiK}.  The result 
is depicted in Fig.\ \ref{fig:fpi}, wherein it is compared with the most 
recent experimental data \cite{Volmer:2000ek}.  It is noteworthy that all 
other pre-existing calculations are uniformly two -- four standard deviations 
below that $Q^2 F_\pi(Q^2)$ data.\footnote{The nature and meaning of vector 
dominance is discussed in Sec.\ 2.3.1 of Ref.\ \protect\cite{cdrpion}, Sec.\ 
2.3 of Ref.\ \protect\cite{bastirev} and Sec.\ 4.3 of Ref.\ 
\protect\cite{pieterrev}: the low-$q^2$ behaviour of the pion form factor is 
necessarily dominated by the lowest mass resonance in the $J^{PC}=1^{--}$ 
channel.  Any realistic calculation will predict that and also a deviation 
from dominance by the $\rho$-meson pole alone as spacelike-$q^2$ increases.} 
 
In this connection one should also note that it is a model independent DSE 
prediction \cite{Maris:1998hc} that electromagnetic elastic meson form 
factors display 
\begin{equation} 
\label{q2Fq2} q^2 F(q^2) = {\rm constant},\; q^2 \gg \Lambda_{\rm QCD}^2, 
\end{equation} 
with calculable $(\ln q^2/\Lambda_{\rm QCD}^2)^d$ corrections, where $d$ is 
an anomalous di\-men\-sion.  This agrees with earlier perturbative QCD 
analyses \cite{Farrar:1979aw,Lepage:1980fj}.  However, to obtain this result 
in covariant gauges it is crucial to retain the pseudovector components of 
the Bethe-Salpeter amplitude in Eq.\ (\ref{genpibsa}): $F_\pi$, $G_\pi$. 
(NB.\ The quark-level Goldberger-Treiman relations, Eqs. (\ref{bfgwti}), 
prove them to be nonzero.)  Without these amplitudes \cite{cdrpion}, $q^2 
F(q^2) \propto 1/q^2$.  The calculation of Ref.\ \cite{Maris:1998hc} suggests 
that the perturbative behaviour of Eq.\ (\ref{q2Fq2}) is unambiguously 
evident for $q^2\gsim 15\,$GeV$^2$.  Owing to challenges in the numerical 
analysis, the \textit{ab initio} calculations of Ref.\ \cite{pieterpiK} 
cannot yet make a prediction for the onset of the perturbative domain but 
progress in remedying that is being made \cite{pichowskypoles}. 
 
Another very instructive success is the study of $\pi$-$\pi$ scattering, 
wherein a range of new challenges arise whose quiddity and natural resolution 
via a symmetry-preserving truncation of the DSEs is explained in Sec.\ 4.6 of 
Ref.\ \cite{pieterrev}, which reviews the seminal work of Refs.\ 
\cite{bicudo}.  It is worth remarking, too, that with a systematic and 
nonperturbative DSE truncation scheme, all consequences of the Abelian 
anomaly and Wess-Zumino term are obtained exactly, without fine tuning 
\cite{cdrpion,Maris:1998hc,WZterm,bando,racdr,Maris:2002mz}. 
 
\subsection{Heavier Mesons} 
The meson spectrum contains \cite{pdg98} four little-studied axial-vector 
mesons composed of $u$- and $d$-quarks.  They appear as isospin $I=0,1$ 
partners (in the manner of the $\omega$ and $\rho$): $h_1(1170)$, 
$b_1(1235)$; and $f_1(1285)$, $a_1(1260)$, and differ in their charge-parity: 
$J^{PC}=1^{+-}$ for $h_1$, $b_1$; and $J^{PC}= 1^{++}$ for $f_1$, $a_1$.  In 
the $q\bar q$ constituent quark model the $b_1$ is represented as a 
constituent-quark and -antiquark with total spin $S=0$ and angular momentum 
$L=1$, while in the $a_1$ the quark and antiquark have $S=1$ and $L=1$.  It 
is therefore apparent that in this model the $b_1$ is an orbital excitation 
of the $\pi$, and the $a_1$ is an orbital excitation and axial-vector partner 
of the $\rho$.  In QCD the $J^{PC}$ characteristics of a quark-antiquark 
bound state are manifest in the structure of its Bethe-Salpeter amplitude 
\cite{llewellyn}.  This amplitude is a valuable intuitive guide and, in cases 
where a $q\bar q$ constituent quark model analogue exists, it incorporates 
and extends the information present in that analogue's quantum mechanical 
wave function. 
 
Three of the axial-vector mesons decay predominantly into two-body final 
states containing a vector meson and a pion: $h_1\to \rho\pi$; $b_1\to \omega 
\pi$; $a_1\to \rho\pi$. 
With a $J=1$ meson in both the initial and final state these three decays 
proceed via two partial waves ($S$, $D$), and therefore probe aspects of 
hadron structure inaccessible in simpler processes involving only spinless 
mesons in the final state, such as $\rho\to\pi\pi$.  For example and of 
importance, in constituent-quark-like models the $D/S$ amplitude ratio is 
very sensitive \cite{eric} to the nature of the phenomenological long-range 
confining interaction. 
 
The additional insight and model constraints that such processes can provide 
is particularly important now as a systematic search and classification of 
``exotic'' states in the light meson sector becomes feasible experimentally. 
I note that a meson is labelled ``exotic'' if it is characterised by a value 
of $J^{PC}$ which is unobtainable in the $q\bar q$ constituent quark model; 
e.g., the experimentally observed \cite{exotic} $\pi_1(1600)$, a $1.6\,$GeV 
$J^{PC}=1^{-+}$ state.  Such unusual charge parity states are a necessary 
feature of a field theoretical description of quark-antiquark bound states 
\cite{llewellyn} with BSE studies typically yielding \cite{bsesep} masses 
approximately twice as large as that of the natural charge parity partner 
and, in particular, a $J^{PC}=1^{-+}$ meson with a mass $\sim 1.5\,$GeV 
\cite{bpprivate}. 
 
In appreciation of these points, Ref.\ \cite{a1b1} used the simple DSE-based 
model of Ref.\ \cite{bsesep} in a simultaneous study of axial-vector meson 
decays, $\rho\to\pi\pi$ decay, and the electroweak decay constants of the 
mesons involved.  The results are instructive.  It was found that the 
rainbow-ladder truncation is capable of simultaneously providing a good 
description of these observables but that the $D/S$ partial-wave ratio in the 
decays of axial-vector mesons is indeed very sensitive to details of the 
long-range part of a model interaction; i.e., to the expression of 
light-quark confinement.  This is perhaps unsurprising given that the mass of 
each axial-vector meson mass is significantly greater than $2 M(0)$; namely, 
twice the constituent-quark mass-scale.  Unfortunately, more sophisticated 
calculations are lacking.  This collection of experimentally well-understood 
mesons has many lessons to teach and should no longer be ignored. 
 
\section{Heavy Quarks} 
\label{sec:HQ} 
\subsection{Features of the Mass Function} 
\label{subsec:HQ} The DSE methods described hitherto have been applied to 
mesons involving heavy-quarks \cite{mishaSVY,misha1,misha2} and in this case 
there is a natural simplification.  To begin, one focuses on the fact that 
mesons, whether heavy or light, are bound states of a dressed-quark and 
-antiquark, with the dressing described by the gap equation, Eq.\ 
(\ref{gendse}), written explicitly again here with the addition of flavour 
label, $f\,(=u,d,s,c,b)$: 
\begin{eqnarray} 
\label{genS} 
\lefteqn{ 
S_f(p)^{-1}  =  i \gamma\cdot p \,A_f(p^2) + B_f(p^2) = A_f(p^2) 
\left[ i \gamma\cdot p + M_f(p^2) \right] 
} \\ 
\label{gendsef} & = & Z_2 (i\gamma\cdot p + m_f^{\rm bm}) 
+\, Z_1 \int^\Lambda_q \!  g^2 D_{\mu\nu}(p-q) \frac{\lambda^a}{2}\gamma_\mu 
S_f(q) \Gamma^{fa}_\nu(q,p) \,. 
\end{eqnarray} 
The other elements of Eq.\ (\ref{gendsef}) will already be familiar. 
 
\begin{figure}[t] 
\centerline{\includegraphics[width=.60\textwidth]{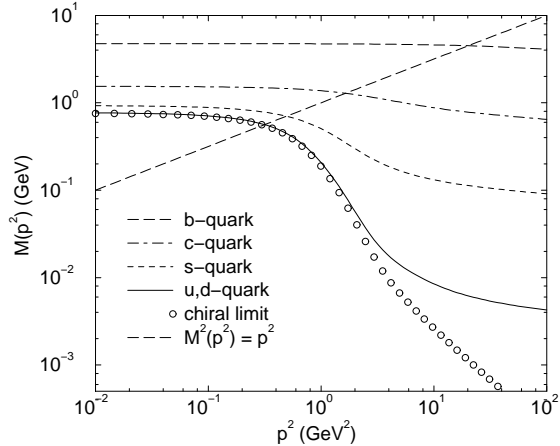}} 
\caption{Quark mass function obtained as a solution of 
Eq.~(\protect\ref{genS}) using the rainbow truncation, discussed in 
connection with Eqs.\ (\protect\ref{RL1}) -- (\protect\ref{rainbowdse}), and 
the interaction of Eq.\ (\protect\ref{gk2}) with current-quark masses, fixed 
at $\zeta= 19\,$GeV: $m_{u,d}(\zeta) = 3.7\,$MeV, $m_s(\zeta) = 82\,$MeV, 
$m_c(\zeta)=0.58\,$GeV and $m_b(\zeta)=3.8\,$GeV. 
The indicated solutions of $M^2(p^2)=p^2$ define a Euclidean 
constituent-quark mass, $M^E_f$, which takes the values: $M^E_u=0.56\,$GeV, 
$M^E_s=0.70\,$GeV, $M^E_c= 1.3\,$GeV, $M^E_b= 4.6\,$GeV. 
\label{mp2fig}} 
\end{figure}

The qualitative features of the gap equation's solution are known and typical 
mass functions, $M_f(p^2)$, are depicted in Fig.\ \ref{mp2fig}.  There is some 
quantitative model-dependence in the momentum-evolution of the mass-function 
into the infrared.  However, with any \textit{Ansatz} for the effective 
interaction that provides an accurate description of $f_{\pi,K}$ and 
$m_{\pi,K}$, one obtains solutions with profiles like those illustrated in the 
figure.  Owing to Eq.\ (\ref{LqkPuv}) the ultraviolet behaviour is naturally 
fixed, namely, it is given by Eq.\ (\ref{hatm}) for massive quarks and by Eq.\ 
(\ref{Mchiral}) in the chiral limit. 
 
It is apparent in the figure that as $p^2$ decreases the chiral-limit and 
$u,d$-quark mass functions evolve to coincidence.  This feature signals a 
transition from the perturbative to the nonperturbative domain.  Furthermore, 
since the chiral limit mass-function is nonzero {\it only} because of the 
nonperturbative DCSB mechanism, whereas the $u,d$-quark mass function is 
purely perturbative at $p^2>20\,$GeV$^2$, it also indicates clearly that the 
DCSB mechanism has a significant impact on the propagation characteristics of 
$u,d,s$-quarks.  However, it is conspicuous in Fig.\ \ref{mp2fig} that this 
is not the case for the $b$-quark.  Its large current-quark mass almost 
entirely suppresses momentum-dependent dressing so that $M_b(p^2)$ is nearly 
constant on a substantial domain.  The same is true to a lesser extent for 
the $c$-quark. 
 
The quantity ${\cal L}_f:=M^E_f/m_f(\zeta)$ provides a single quantitative 
measure of the importance of the DCSB mechanism; i.e., nonperturbative 
effects, in modifying the propagation characteristics of a given quark 
flavour.  In this particular illustration it takes the values 
\begin{equation} 
\label{Mmratio} 
\setlength\arraycolsep{5pt} 
\begin{array}{c|ccccc} 
        f   &   u,d  &   s   &  c  &  b  \\ \hline 
 {\cal L}_f &  150   &    10      &  2.2 &  1.2 
\end{array}\,, 
\end{equation} 
which are representative: for light-quarks ${\cal L}_{q=u,d,s} \sim 
10$-$100$; while for heavy-quarks ${\cal L}_{Q=c,b} \sim 1$.  They also 
highlight the existence of a mass-scale, $M_\chi$, characteristic of DCSB: 
the propagation characteristics of a flavour with $m_f(\zeta)\leq M_\chi$ are 
significantly altered by the DCSB mechanism, while momentum-dependent 
dressing is almost irrelevant for flavours with $m_f(\zeta)\gg M_\chi$.  It 
is evident and unsurprising that $M_\chi \sim 0.2\,$GeV$\,\sim \Lambda_{\rm 
QCD}$.  Consequently one anticipates that the propagation of $c,b$-quarks 
should be described well by replacing their mass-functions with a constant; 
viz., writing \cite{mishaSVY} 
\begin{equation} 
\label{dsehq} 
S_Q(p) = \frac{1}{i\gamma \cdot p + \hat M_Q}\,,\; Q=c,b\,, 
\end{equation} 
where $\hat M_Q$ is a constituent-heavy-quark mass 
parameter.\footnote{Although not illustrated explicitly, when $M_f(p^2) 
\approx\,$const., $A_f(p^2)\approx 1$ in Eq.\ (\protect\ref{genS}).} 
 
When considering a meson with an heavy-quark constituent one can proceed 
further, as in heavy-quark effective theory (HQET) \cite{neubert94}, allow the 
heaviest quark to carry all the heavy-meson's momentum: $ P_\mu=: m_H v_\mu=: 
(\hat M_Q + E_H)v_\mu $, and write 
\begin{equation} 
\label{hqf} 
S_Q(k+P) = \frac{1}{2}\,\frac{1 - i \gamma\cdot v}{k\cdot v - E_H} 
+ {\rm O}\left(\frac{|k|}{\hat M_{Q}}, 
                \frac{E_H}{\hat M_{Q}}\right)\,, 
\end{equation} 
where $k$ is the momentum of the lighter constituent.  It is apparent from 
the study of light-meson properties that in the calculation of observables 
the meson's Bethe-Salpeter amplitude will limit the range of $|k|$ so that 
Eq.\ (\ref{hqf}) will only be a good approximation if {\it both} the 
momentum-space width of the amplitude, $\omega_H$, and the binding energy, 
$E_H$, are significantly less than $\hat M_Q$. 
 
In Ref.\ \cite{misha2} the propagation of $c$- and $b$-quarks was described by 
Eq.\ (\ref{hqf}), with a goal of exploring the fidelity of this idealisation, 
and it was found to allow for a uniformly good description of $B_f$-meson 
leptonic and semileptonic decays with heavy- and light-pseudoscalar final 
states.  In that study, $\omega_{B_f} \approx 1.3\,$GeV and $E_{B_f} \approx 
0.70\,$GeV, both of which are small compared with $\hat M_b \approx 4.6\,$GeV 
in Fig.\ \ref{mp2fig}.  Hence the accuracy of the approximation could be 
forseen. It is reasonable to suppose that $\omega_D\approx \omega_B$ and 
$E_D\approx E_B$, since they must be identical in the limit of exact 
heavy-quark symmetry. Thus in processes involving the weak decay of a $c$-quark 
($\hat M_c\approx 1.3\,$GeV) where a $D_f$-meson is the heaviest participant, 
Eq.\ (\ref{hqf}) must be inadequate; an expectation verified in 
Ref.~\cite{misha2}. 
 
The failure of Eq.~(\ref{hqf}) for the $c$-quark complicates or precludes the 
development of a common understanding of $D_f$- and $B_f$-meson observables 
using such contemporary theoretical tools as HQET and light cone sum rules. 
However, as shown in Ref.\ \cite{mishaSVY} and I will illustrate, the 
constituent-like dressed-heavy-quark propagator of Eq.\ (\ref{dsehq}) can still 
be used to effect a unified, accurate simplification in the study of these 
observables. 
 
\subsection{Leptonic Decays} 
\subsubsection{Pseudoscalar Mesons.}  The leptonic decay of a pseudoscalar 
meson, $P(p)$, is described by the matrix element (Sec.\ 
\ref{sec:massformula}) 
\begin{eqnarray} 
\label{fwk} 
f_{P} \,p_\mu := \langle 0| \bar {\cal Q}\, (T^P)^{\rm T} 
\gamma_\mu \gamma_5 \,{\cal Q} | P(p) \rangle 
= {\rm tr}\,Z_2 \int^\Lambda_k \,(T^P)^{\rm T}\gamma_5\gamma_\mu\, 
        \chi_P(k;p), 
\end{eqnarray} 
where ${\cal Q}= {\rm column}(u,d,s,c,b)$ and here I have adopted a charged 
particle normalisation, which yields results for $f_P$ a factor of $\surd 2$ 
larger than Eq.\ (\protect\ref{fH}) and is conventional in studying heavy-quark 
systems. 
 
In Eq.\ (\ref{fwk}), $\chi_P$ is the meson's Bethe-Salpeter wave function, 
related to its amplitude, $\Gamma_P$, via Eq.\ (\ref{chiM}) and normalised 
canonically as described in connection with Eq.\ (\ref{bsegen}).  Using Eq.\ 
(\ref{hqf}), it follows from the canonical normalisation condition that 
\begin{equation} 
\label{calGP} 
{\cal G}_P(k;p) := \frac{1}{\surd m_P}\, 
\Gamma_P(k;p) < \infty\,, \; m_P \to \infty\,; 
\end{equation} 
i.e., ${\cal G}_P(k;p)$ so-defined is mass-independent in the heavy-quark 
limit.  Using this result plus Eq.\ (\ref{hqf}) one finds from Eq.\ (\ref{fwk}) 
\cite{misha1} 
\begin{equation} 
\label{fwkasymp} 
f_P \propto \frac{1}{\surd{m_P}}\,, \; m_P \to \infty \,. 
\end{equation} 
 
\begin{figure}[t] 
\centerline{\includegraphics[width=.60\textwidth]{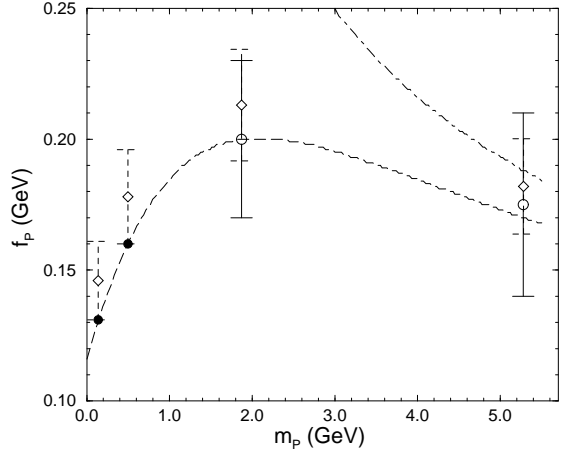}} 
\caption{\label{pic:fWK} Experimental values of $f_{\pi,K}$ (Eq.\ 
(\protect\ref{fpifK}), \textit{filled circles}); lattice estimates of 
$f_{D,B}$ (Eq.~(\protect\ref{fDfB}), \textit{open circles}); values of 
$f_{\pi,K,D,B}$ calculated in Ref.\ \protect\cite{mishaSVY} (\textit{open 
diamonds}).  Least-squares fit to the experimental values and lattice 
estimates (\textit{dashed curve}): \mbox{$f_P^2= (0.013 + 0.028 \,m_P )/(1+ 
0.055\,m_P + 0.15 \,m_P^2)$}, which exhibits the large-$m_P$ limit of 
Eq.~(\protect\ref{fwkasymp}); the large-$m_P$ limit of this fit 
(\textit{dot-dashed curve}).  (Adapted from Ref.\ \protect\cite{mishaSVY}.)} 
\end{figure}

Equation (\ref{fwkasymp}) is a model-independent result and a well-known 
general consequence of heavy-quark symmetry \cite{neubert94}.  However, the 
value of the hadron mass at which this behaviour becomes evident is unknown. 
It is clear from Table \ref{RGIpiK} that, experimentally, 
\begin{equation} 
\label{fpifK} 
f_\pi = 131\, {\rm MeV}\; < \; f_K = 160\, {\rm MeV}. 
\end{equation} 
Furthermore, direct DSE studies following the method described in Sec.\ 
\ref{abinitio} show that for pseudoscalar mesons $u\bar f$, composed of a 
single $u,d$-quark and an antiquark of mass $m_f$, $f_P(m_P)$ is a 
monotonically increasing concave-down function on $m_P \in [0,0.9]\,{\rm GeV}$, 
where $m_P$ is the calculated mass of this composite system, and likely on a 
larger domain \cite{TandyErice}.  On the other hand, numerical simulations of 
quenched lattice-QCD indicate \cite{flynn} 
\begin{equation} 
\label{fDfB} 
f_D = 200 \pm 30 \,{\rm MeV}\; > \; f_B = 170 \pm 35\, {\rm MeV}\,. 
\end{equation} 
In simulations of lattice-QCD with two flavours of sea quarks both of these 
decay constants increase in magnitude but there is no sign that the ordering 
is reversed \cite{flynn,mcneile}.  The information in Eqs.\ (\ref{fpifK}), 
(\ref{fDfB}) is depicted in Fig.\ \ref{pic:fWK}.  This and analysis to be 
reviewed subsequently suggest that D-mesons lie outside the domain on which 
Eq.~(\protect\ref{fwkasymp}) is a reliable tool. 
 
\subsubsection{Vector Mesons.} 
The leptonic decay constant, $f_V$, for a vector meson with mass $M_V$ is 
given in Eq.\ (\ref{fV}) and adapting the analysis that leads to Eq.\ 
(\ref{fwkasymp}) one finds readily 
\begin{equation} 
f_V \propto \frac{1}{\sqrt{M_V}}\,, \; M_V \to \infty \,, 
\end{equation} 
which again is a model-independent result.  Moreover, since the pseudoscalar 
and vector meson Bethe-Salpeter amplitudes become identical in the 
heavy-quark limit, it follows that \cite{mishaSVY} 
\begin{equation} 
f_V = f_P\,, \; M_V = m_P \,,\; \mbox{\rm in the limit} \; m_P\to \infty\,. 
\end{equation} 
 
\subsection{Heavy-Meson Masses} 
More can be learnt from the pseudoscalar meson mass formula in Eq.\ 
(\ref{massformula}).  Using Eq.\ (\ref{fwkasymp}), and applying to Eq.\ 
(\ref{qbqH}) the analysis from which it follows, one obtains 
\begin{equation} 
-\langle \bar q q \rangle^P_\zeta = \mbox{constant}\,,\; \mbox{as~} m_P \to 
 \infty 
\end{equation} 
and consequently \cite{marisAdelaide,mishaSVY} 
\begin{equation} 
m_P \propto \hat m_Q\,, \; \hat m_Q \to \infty\,, 
\end{equation} 
where $\hat m_Q$ is the renormalisation-group-invariant current-quark mass of 
the fla\-vour-nonsinglet pseudoscalar meson's heaviest constituent.  This is 
the result one would have guessed from constituent-quark models but here I 
have outlined a direct proof in QCD. 
 
\begin{figure}[t] 
\centerline{\includegraphics[width=.60\textwidth]{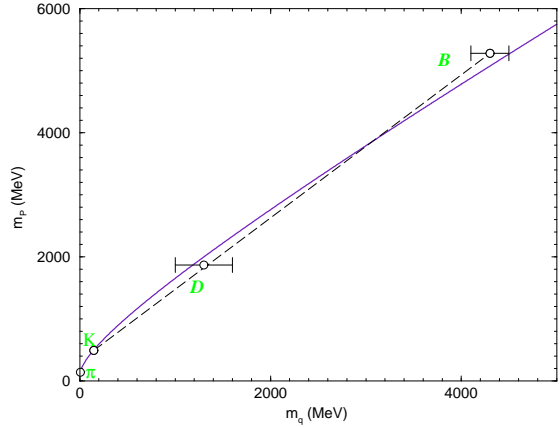}} 
\caption{\label{figmH} Pseudoscalar $u\bar q$ meson's mass as a function of 
$m_q({\zeta})$, $\zeta=19\,{\rm GeV}$, with a fixed value of $m_u({\zeta})$ 
corresponding to $m_u(1\,{\rm GeV})=5.5\,$MeV, Eq.\ (\protect\ref{mqs}) 
(\textit{solid line}).  The experimental data points are from Ref.\ 
\protect\cite{pdg98} as are the errors assigned to the associated heavy-quark 
masses.  A straight line is drawn through the $K$, $D$, $B$ masses 
(\textit{dashed curve}).  (Adapted from Ref.\ \protect\cite{cdrlc01}.  See also 
Ref.\ \protect\cite{TandyErice}.)  } 
\end{figure} 
 
Equation (\ref{gmora}) is thus seen to be a single formula that unifies the 
masses of light- and heavy-quark mesons.  This aspect has been quantitatively 
explored using the rainbow-ladder kernel described in Sec.\ \ref{sect4label}, 
with the results illustrated in Fig.~\ref{figmH}.  Therein the calculated mass 
of a $u \bar f$ pseudoscalar meson is plotted as a function of $m_f(\zeta)$, 
with $m_u(\zeta)$ fixed at the value in Eq.\ (\ref{mqs}).  The DSE calculations 
are depicted by the solid curve, which is \cite{pmqciv} (in MeV) 
\begin{equation} 
\label{pietermH} m_P = 83 + 500 \sqrt{\cal X} + 310\,{\cal X},\; 
{\cal X}= m_q^{\zeta}/\Lambda_{\rm QCD}. 
\end{equation} 
The curvature appears slight in the figure but that is misleading: the
nonlinear term in Eq.\ (\ref{pietermH}) accounts for almost all of $m_\pi$
(the Gell-Mann--Oakes-Renner relation is nearly exact for the pion) and
$80\,$\% of $m_K$.  NB.\ The dashed line in Fig.\ \ref{figmH} fits the $K$,
$D$, $B$ subset of the data exactly.  It is drawn to illustrate how easily
one can be misled.  Without careful calculation one might infer from this
apparent agreement that the large-$m_q$ limit of Eq.\ (\ref{gmora}) is
already manifest at the $s$-quark mass whereas, in reality, the linear term
only becomes dominant for $m_q \gsim 1\,$GeV, providing $50\,$\% of $m_D$ and
$67\,$\% of $m_B$.  The model predicts, via Eq.\ (\ref{Zmone}), $m_c^{1\,{\rm
GeV}}=1.1\,$GeV and $m_b^{1\,{\rm GeV}}=4.2\,$GeV, values that are typical of
Poincar\'e covariant treatments.
 
\subsection{Semileptonic Transition Form Factors} 
\label{secd} 
\subsubsection{Pseudoscalar meson in the final state.} 
The transition: $P_1(p_1) \to P_2(p_2)\,\ell\,\nu\,,$ where $P_1$ represents 
either a $B$- or $D$-meson and $P_2$ can be a $D$, $K$ or $\pi$, is described 
by the invariant amplitude 
\begin{equation} 
A(P_1 \to P_2\,\ell\,\nu) 
= \frac{G_F}{\surd 2}\, V_{f^\prime f}\; 
\bar \ell \,\gamma_\mu(1 - \gamma_5)\nu\; 
M_\mu^{P_1 P_2}(p_1,p_2)\,, 
\end{equation} 
where $G_F=1.166 \times 10^{-5}\,$GeV$^{-2}$, $V_{f^\prime f}$ is the relevant 
element of the Cabibbo-Kobayashi-Maskawa (CKM) matrix, and the hadronic 
current is 
\begin{eqnarray} 
\label{fpfm} 
M_\mu^{P_{1} P_{2}}(p_1,p_2)  & := & 
\langle P_{2}(p_2)| \bar f^\prime \gamma_\mu f | P_{1}(p_1)\rangle 
 =  f_+(t)\, (p_1 + p_2)_\mu + f_-(t) \,q_\mu\,, 
\end{eqnarray} 
with $t:= -q^2 = -(p_1-p_2)^2$.  The transition form factors, $f_\pm(t)$, 
contain all the information about strong-interaction effects in these 
processes, and their accurate calculation is essential for a reliable 
determination of the CKM matrix elements from a measurement of the decay 
width ($t_\pm := (m_{P_1}\pm m_{P_2})^2$): 
\begin{equation} 
\label{branching} 
\Gamma(P_{1} \to P_{2}\ell\nu)= 
\frac{G_F^2}{192 \pi^3}\,|V_{f^\prime f}|^2\,\frac{1}{m_{P_1}^3}\, 
\int_0^{t_-}\,dt\,|f_+(t)|^2\, 
\left[(t_+-t) (t_- - t)\right]^{3/2}\!\!. 
\end{equation} 
The related study of light-meson initial states is described in Refs.\ 
\cite{mitchell}. 
 
\subsubsection{Vector meson in the final state.} 
The transition: $P(p_1) \to V_\lambda(p_2)\,\ell\,\nu\,,$ with $P$ either a 
$B$ or $D$ and $V_\lambda$ a $D^\ast$, $K^\ast$ or $\rho$, is described by 
the invariant amplitude 
\begin{equation} 
A(P \to V_\lambda\,\ell\,\nu) 
= \frac{G_F}{\surd 2}\, V_{f^\prime f}\, 
\bar \ell \,i\gamma_\mu(1 - \gamma_5)\nu\,\epsilon_\nu^\lambda(p_2)\, 
M_{\mu\nu}^{P V_\lambda}(p_1,p_2)\,, 
\end{equation} 
in which the hadronic tensor involves four scalar functions 
\begin{eqnarray} 
\nonumber \epsilon_\nu^\lambda(p_2)\,M_{\mu\nu}^{P V_\lambda}(p_1,p_2) & = & 
\epsilon_\mu^\lambda\,(m_P + M_V)\,A_1(t) + (p_1+p_2)_\mu\,\epsilon^\lambda 
\cdot q\,\frac{A_2(t)}{m_P + M_V}\\ 
\nonumber &+ & 
  q_\mu\,\epsilon^\lambda \cdot q\,\frac{A_3(t)}{m_P + M_V}\, 
+\, \varepsilon_{\mu\nu\alpha\beta}\, 
        \epsilon_\nu^\lambda \,p_{1\alpha}\, p_{2\beta} \, 
        \frac{2 V(t)}{m_P + M_V}\,.\\ 
\label{vsl} 
\end{eqnarray} 
Introducing three helicity amplitudes 
\begin{eqnarray} 
H_\pm & = & (m_P + M_V)\, A_1(t)\, \mp \, 
\frac{\lambda^{\frac{1}{2}}(m_P^2,M_V^2,t)}{m_P + M_V}\, V(t)\,,\\ 
\nonumber H_0  & = & \frac{1}{2\,M_V\surd t}\left( 
        \left[ m_P^2 - M_V^2 -t\right] \, \left[ m_P + M_V\right] A_1(t) 
        - \frac{\lambda(m_P^2,M_V^2,t)}{m_P + M_V}\,A_2(t)\right)\!,\\ 
\end{eqnarray} 
where $\lambda(m_P^2,M_V^2,t)= [t_+ -t]\, [t_- -t]$, $t_\pm = (m_P\pm 
M_V)^2$, the transition rates 
\begin{eqnarray} 
\frac{d\Gamma_{\pm,0}}{dt} & = & 
\frac{G_F^2}{192 \pi^3 m_{P_1}^3}\,|V_{f^\prime f}|^2\,t\, 
\lambda^{\frac{1}{2}}(m_P^2,M_V^2,t)\,|H_{\pm,0}(t)|^2\,, 
\end{eqnarray} 
from which one obtains the transverse and longitudinal rates
\begin{eqnarray} 
\frac{d\Gamma_{T}}{dt} & = & 
        \frac{d\Gamma_{+}}{dt} + \frac{d\Gamma_{-}}{dt}\,,\; 
\Gamma_T= \int_0^{t_-}\,dt\,\frac{d\Gamma_T}{dt}\,, 
\\ 
\frac{d\Gamma_{L}}{dt} & = & \frac{d\Gamma_{0}}{dt}\,,\; 
\Gamma_L= \int_0^{t_-}\,dt\,\frac{d\Gamma_L}{dt}\,, 
\end{eqnarray} 
wherefrom the total width $\Gamma= \Gamma_T + \Gamma_L$.  The polarisation
ratio and forward-backward asymmetry are
\begin{eqnarray} 
\alpha = 2 \,\frac{\Gamma_L}{\Gamma_T} - 1\,, &\;& 
A_{\rm FB} = \frac{3}{4}\,\frac{\Gamma_- - \Gamma_+}{\Gamma}\,. 
\end{eqnarray} 
 
\subsection{Impulse Approximation} 
As explained and illustrated in Ref.\ \cite{pieterrev}, the impulse 
approximation is accurate for three point functions and applied to these 
transition form factors it yields 
\begin{eqnarray} 
\nonumber \lefteqn{{\cal H}^{PX}_\mu(p_1,p_2) = }\\ && 2 N_c\,{\rm 
tr}_D\,\int_k^\Lambda\, \bar\Gamma_X(k;-p_2)\,S_{q}(k_2) \, i{\cal 
O}_\mu^{qQ}(k_2,k_1)\, S_Q(k_1)\,\Gamma_P(k;p_1)\,S_{q^\prime}(k), 
\label{ia} 
\end{eqnarray} 
wherein the flavour structure is made explicit, $k_{1,2} = k+p_{1,2}$ and: 
\begin{eqnarray} 
{\cal H}^{P_1 X=P_2}_\mu(p_1,p_2) &=& M_\mu^{P_1 P_2}(p_1,p_2)\,,\\ 
{\cal H}^{P X=V^\lambda}_\mu(p_1,p_2) & = & 
\epsilon_\nu^\lambda(p_2)\,M_{\mu\nu}^{P V_\lambda}(p_1,p_2)\,; 
\end{eqnarray} 
$\Gamma_{X=V^\lambda}(k;p)= \epsilon^\lambda(p)\cdot \Gamma^V(k;p)$; and 
${\cal O}_\mu^{qQ}(k_2,k_1)$ is the dressed-quark-W-boson vertex, which in 
weak decays of heavy-quarks is well approximated by \cite{misha1,misha2} 
\begin{equation} 
{\cal O}_\mu^{qQ}(k_2,k_1) = \gamma_\mu\,(1 - \gamma_5) 
\end{equation} 
because $A_Q(p^2)\approx\,$const.\ and $M_Q(p^2)\approx\,$const.\ for 
heavy-quarks (recall Fig.\ \ref{mp2fig}.) 
 
\subsubsection{Quark Propagators.} It is plain that  to evaluate ${\cal H}^{P_1 
X}_\mu(p_1,p_2)$ a specific form for the dressed-quark propagators is 
required.  Equation (\ref{dsehq}) provides a good approximation for the 
heavier quarks, $Q=c,b$, as explained in Sec.\ \ref{subsec:HQ}, and this was 
used in Ref.\ \cite{mishaSVY} with $\hat M_Q$ treated as free parameters. 
 
For the light-quark propagators: 
\begin{equation} 
\label{defS} 
S_f(p) = -i\gamma\cdot p\,\sigma^f_V(p^2) + \sigma^f_S(p^2) 
        = \frac{1}{i\gamma\cdot p\,A_f(p^2) + B_f(p^2)}\,, 
\end{equation} 
Ref.\ \cite{mishaSVY} assumed isospin symmetry and employed the algebraic forms 
introduced in Ref.~\cite{cdrpion}, which efficiently characterise the essential 
features of the gap equation's solutions: 
\begin{eqnarray} 
\label{SSM} 
\bar\sigma^f_S(x)  & =  & 
        2 \bar m_f {\cal F}(2 (x + \bar m_f^2)) 
        + {\cal F}(b_1 x) {\cal F}(b_3 x) 
        \left[ b^f_0 + b^f_2 {\cal F}(\epsilon x)\right],\\ 
\label{SVM} 
\bar\sigma^f_V(x) & = & \frac{2 (x+\bar m_f^2) -1 
                + e^{-2 (x+\bar m_f^2)}}{2 (x+\bar m_f^2)^2}\,, 
\end{eqnarray} 
${\cal F}(y)= (1-{\rm e}^{-y})/y$, $x=p^2/\lambda^2$; $\bar m_f$ = 
$m_f/\lambda$; and $\bar\sigma_S^f(x) = \lambda\,\sigma_S^f(p^2)$, 
$\bar\sigma_V^f(x) = \lambda^2\,\sigma_V^f(p^2)$, with $\lambda$ a mass 
scale.  The parameters are the current-quark mass, $\bar m$, and 
$b_{0,1,2,3}$, about which I shall subsequently explain more. 
 
This algebraic form combines the effects of confinement\footnote{The 
representation of $S(p)$ as an entire function is motivated by the algebraic 
solutions of Eq.~(\protect\ref{gendse}) in Refs.~\cite{munczekburden}.  The 
concomitant violation of the axiom of \textit{reflection positivity} is a 
sufficient condition for confinement, as reviewed in Sec.\ 6.2 of Ref.\ 
\cite{cdragw}, Sec.\ 2.2 of Ref.\ \cite{bastirev} and Sec.\ 2.4 of Ref.\ 
\cite{reinhardrev}.}\  
and DCSB with free-particle behaviour at large, spacelike $p^2$.  One 
characteristic of DCSB is the appearance of a nonzero vacuum quark condensate 
and using this parametrisation in Eq.\ (\ref{qbq0}) yields 
\begin{eqnarray} 
-\langle \bar u u \rangle_\zeta & = & \lambda^3\,\frac{3}{4\pi^2}\, 
\frac{b_0^u}{b_1^u\,b_3^u}\,\ln\frac{\zeta^2}{\Lambda_{\rm QCD}^2}\,. 
\end{eqnarray} 
The simplicity of this result emphasises the utility of an algebraic form for 
the dressed-quark propagator.  That utility is amplified in the calculation 
of a form factor, which requires the repeated evaluation of a 
multidimensional integral whose integrand is a complex-valued function, and a 
functional of the propagator and the Bethe-Salpeter amplitudes. 
 
\subsubsection{Bethe-Salpeter Amplitudes.} 
An algebraic parametrisation of the Bethe-Salpeter amplitudes also helps and 
the quark-level Goldberger-Treiman relation, Eq.\ (\ref{bfgwti}), suggests a 
form for light pseudoscalar mesons: 
\begin{equation} 
\label{piKamp} 
\Gamma_{P}(k;p)= i \gamma_5 \,{\cal E}_{P}(k^2) = i \gamma_5 \,\frac{1}{\hat 
f_{P}}\,B_{P}(k^2)\,,\;P=\pi,K\,, 
\end{equation} 
where $B_P:=\left.B_u\right|_{b_0^u\to b_0^P}$, obtained from Eq.~(\ref{defS}), 
and $\hat f_P= f_P/\surd 2$ because in this Section I use the $f_\pi = 
131\,$MeV normalisation.  $b_0^{\pi,K}$ are two additional parameters.  This 
\textit{Ansatz} omits the pseudovector components of the amplitude but that is 
not a material defect in applications involving small to intermediate momentum 
transfers \cite{Maris:1998hc}.  Equations (\ref{massformula}), (\ref{qbqH}), 
(\ref{piKamp}) yield the following expression for the $\pi$- and $K$-meson 
masses: 
\begin{equation} 
\hat f_{P}^2 \, m_P^2 = -( m_u +  m_{f^P})\, 
\langle \bar q q\rangle^P_{1\,{\rm GeV}^2}\,, 
\end{equation} 
where $m_{f^\pi}=m_d$, $m_{f^K}=m_s$, and 
\begin{equation} 
-\,\langle \bar q q\rangle^P_{1\,{\rm GeV}^2}= 
\lambda^3\,\ln\frac{1}{\Lambda_{\rm QCD}^2}\,\frac{3}{4\pi^2}\, 
\frac{b_0^P}{b_1^u\,b_3^u}\,,\;P=\pi,K\,. 
\end{equation} 
In studies of the type reviewed in Sec.\ \ref{sect4label}, this in-hadron 
condensate takes values $\langle \bar q q\rangle^\pi_{1\,{\rm GeV}^2} \approx 
1.05\,\langle \bar u u \rangle_{1\,{\rm GeV}^2}$ 
and 
$\langle \bar q q\rangle^K_{1\,{\rm GeV}^2} \approx 
1.6\,\langle \bar u u \rangle_{1\,{\rm GeV}^2}$. 
 
Employing algebraic parametrisations of the light vector meson Bethe-Salpeter
amplitudes is also a useful expedient and that approach was adopted in Refs.\
\cite{mishaSVY,misha1,misha2}.  Indeed, sophisticated calculations of light
vector meson properties based on the rainbow-ladder truncation did not exist
at the time of those studies, although it was clear that a given vector meson
is narrower in momentum space than its pseudoscalar partner, and that for
both vector and pseudoscalar mesons this width increases with the total
current-mass of the constituents.  These qualitative features were important
in the explanation of meson electroproduction cross sections \cite{pichowsky}
and electromagnetic form factors \cite{ph98}, and can be realised in the
simple expression
\begin{equation} 
\label{gammaV} 
\Gamma^V_{\mu}(k;p) = \frac{1}{{\cal N}^V}\, 
\left(\gamma_\mu + p_\mu\,\frac{\gamma\cdot p}{M_V^2}\right)\, 
\varphi(k^2)\,, 
\end{equation} 
where $\varphi(k^2) = 1/(1+k^4/\omega_V^4)$ with $\omega_V$ a parameter and
${\cal N}^V$ fixed by the canonical normalisation condition.  One expects:
$\omega_{K^\ast} \approx 1.6\,\omega_\rho$ \cite{ph98}.
 
In connection with the impulse approximation to semileptonic transition form 
factors it remains only to fix the heavy-meson Bethe-Salpeter amplitudes.  In 
this case, too, algebraic parametrisations offer a simple, attractive and 
expeditious means of proceeding and that again was the approach adopted in 
Ref.\ \cite{mishaSVY}.  Therein heavy vector mesons were described by Eq.\ 
(\ref{gammaV}), with $\varphi(k^2)\to \varphi_H(k^2)$, and heavy pseudoscalar 
mesons by its analogue: 
\begin{equation} 
\label{HPamp} 
\Gamma_P(k;p)= \frac{1}{{\cal N}^P}\,i\,\gamma_5\, 
\varphi_H(k^2)\,, 
\end{equation} 
where $\varphi_H(k^2) = \exp\left(-k^2/\omega_H^2\right)$.  The amplitudes 
are again normalised canonically.  Such a parametrisation naturally 
introduces additional parameters; viz., the widths.  The number is kept at 
two by acknowledging that Bethe-Salpeter amplitudes for truly heavy-mesons 
must be spin- and flavour-independent and assuming therefore that $\omega_B = 
\omega_{B^\ast}= \omega_{B_s}$ and $\omega_D = \omega_{D^\ast}= 
\omega_{D_s}$. 
 
\subsection{Additional Decay Processes} 
Many more decays were considered in Ref.\ \cite{mishaSVY}, with the goal
being to determine whether a unified description of light- and heavy-meson
observables is possible based simply on the key DSE features of quark
dressing and sensible bound state amplitudes.  For example, there are
experimental constraints on radiative decays $H^\ast \to H\,\gamma\,,$ where
$H= D_{(s)}$, $B_{(s)}$, and so these widths, $\Gamma_{H^\ast \to H\gamma}$,
were calculated.  The strong decays $H^\ast \to H\, \pi$ were also studied.
They can be characterised by a coupling constant $g_{H^\ast H \pi}$, which is
calculable even if the process is kinematically forbidden, as is $B^\ast \to
B \pi$.  Lastly, the width for the rare flavour-changing neutral current
process $B\to K^\ast \gamma$, which proceeds predominantly via the local
magnetic penguin operator \cite{buchalla} and can be characterised by a
coupling $g_{BK^\ast \gamma}$, was calculated because data exists and this
process might be expected to severely test the framework since it completely
exceeds the scope of previous applications.
 
\subsection{Heavy-Quark Symmetry Limits} 
\label{sec:HQSym} 
Equation (\ref{hqf}), and Eq.\ (\ref{calGP}) and its natural analogues, can 
be used to elucidate the heavy-quark symmetry limit of the impulse 
approximation to any process and many were made explicit in Refs.\ 
\cite{mishaSVY,misha1,misha2}.  I will only recapitulate on the most 
straightforward three-point case; namely, the semileptonic heavy $\to$ heavy 
transitions.  From Eqs.\ (\ref{fpfm}), (\ref{ia}) one obtains 
\begin{equation} 
\begin{array}{ll} 
\displaystyle 
f_\pm(t)= {\cal T}_\pm \,\xi_f(w) 
:= \frac{m_{P_2} \pm 
m_{P_1}}{2\sqrt{m_{P_2}m_{P_1}}}\,\xi_f(w)\,, \\ 
\\ 
\displaystyle 
\xi_f(w)  = 
\kappa_f^2\,\frac{N_c}{4\pi^2}\, 
\int_0^1\! d\tau\,\frac{1}{W}\, 
\int_0^\infty \!\! du \, \varphi_H(z_W)^2\, 
        \left[\sigma_S^f(z_W) + \sqrt{\frac{u}{W}} \sigma_V^f(z_W)\right],\\ 
\end{array} 
\label{xif} 
\end{equation} 
where: $W= 1 + 2 \tau (1-\tau) (w-1)$, $z_W= u - 2 E_H \sqrt{u/W}$; 
\begin{equation} 
\label{kappaf} 
\frac{1}{m_H}\,\frac{1}{\kappa_f^2} := 
 {\cal N}_P^2 = {\cal N}_V^2  = \frac{1}{m_H} \,\frac{N_c}{4\pi^2}\, 
\int_0^\infty\,du\,\varphi^2_H(z)\, 
\left\{ \sigma_S^f(z) + \surd{u}\,\sigma_V^f(z)\right\}\,, 
\end{equation} 
with $z=u-2 E_H \surd u$, $f$ labelling the meson's lighter quark and all 
dimensioned quantities expressed in units of the mass-scale, $\lambda$; and 
\begin{equation} 
\label{defomega} 
w = \frac{m_{P_1}^2 + m_{P_2}^2 - t}{2 m_{P_1} m_{P_2}} = - v_{P_1} \cdot 
v_{P_2}\,. 
\end{equation} 
The canonical normalisation of the Bethe-Salpeter amplitude automatically 
ensures 
\begin{equation} 
\label{xione} 
\xi_f(w=1) = 1 
\end{equation} 
and from Eq.~(\ref{xif}) follows \cite{misha1} 
\begin{equation} 
\rho^2 := -\left.\frac{d\xi_f}{dw}\right|_{w=1} \geq \frac{1}{3}\,. 
\end{equation} 
 
Semileptonic transitions with heavy vector mesons in the final state, described 
by Eqs.\ (\ref{vsl}) and (\ref{ia}), can be analysed in the same way, and that 
yields 
\begin{equation} 
\label{a1xi} 
A_1(t)= \frac{1}{{\cal T}_+}\,\sfrac{1}{2}\,(1+w)\,\xi_f(w) \,,\; 
A_2(t) = -A_3(t) = V(t) = {\cal T}_+\,\xi_f(w)\,. 
\end{equation} 
 
Equations (\ref{xif}), (\ref{a1xi}) are exemplars of a general result that in 
the heavy-quark symmetry limit the semileptonic $H_f \to H_f^\prime$ 
transitions are described by a single, universal function: $\xi_f(w)$ 
\cite{IW90}.  In this limit the functions 
\begin{equation} 
\label{Rratios} 
R_1(w) :=  (1 - t/t_+) \,\frac{V(t)}{A_1(t)}\,,\; 
R_2(w) :=  (1 - t/t_+) \,\frac{A_2(t)}{A_1(t)} 
\end{equation} 
are constant ($=1$), independent of $w$. 
 
\subsection{Survey of Results for Light- and Heavy-Meson Observables} 
With every necessary element defined, the calculation of observables is a 
straightforward numerical exercise.  The algebraic \textit{Ans\"atze} 
described above involve ten parameters plus four current-quark masses and in 
Ref.\ \cite{mishaSVY} they were fixed via a $\chi^2$-fit to the $N_{\rm 
obs}=42$ heavy- and light-meson observables in Table \ref{tableC}, a process 
which yielded \cite{fn:foot} 
\begin{equation} 
\label{tableB} 
\begin{array}{c|lll} 
      & \;\bar m_f& b_1^f & b_2^f\\\hline 
 u\;  & \;0.00948 & 2.94 & 0.733 \\ 
 s\;  & \;0.210   & 3.18 & 0.858 
\end{array}\rule{1.3em}{0ex}\;\; 
\begin{array}{c|l} 
    & b_0^P \\ \hline 
\pi & 0.204 \\ 
K   & 0.319 
\end{array} \rule{1.3em}{0ex} 
\begin{array}{c|l} 
      & \omega_V^{\rm GeV} \\ \hline 
\rho  & 0.515 \\ 
K^\ast& 0.817 
\end{array} \rule{1.3em}{0ex} 
\begin{array}{c|l} 
      & \omega_H^{\rm GeV} \\ \hline 
   D  & 1.81 \\ 
   B  & 1.81 
\end{array}\rule{1.3em}{0ex} 
\begin{array}{c|l} 
      & \hat M_Q^{\rm GeV} \\ \hline 
   c  & 1.32 \\ 
   b  & 4.65 
\end{array} 
\end{equation} 
with $\chi^2/{\rm d.o.f} = 1.75$ and $\chi^2/N_{\rm obs} = 1.17$.  The
dimensionless $u,s$ current-quark masses correspond to $m_u = 5.4\,$MeV,
$m_s=119\,$MeV, and $M^E_u= 0.36 \,$GeV, $M^E_s= 0.49\,$GeV.  Furthermore,
$\omega_{K^\ast}/\omega_\rho=1.59$, which is identical to the value in Ref.\
\cite{ph98}.
 
\begin{table}[t] 
\caption{\label{tableC} The 16 dimension-GeV (\textit{upper panel}) and 26 
dimensionless (\textit{lower panel}) quantities used in the $\chi^2$-fit of 
Ref.\ \protect\cite{mishaSVY}.  The weighting error was the experimental 
error or 10\% of the experimental value, if that is greater, which accounts 
for a realistic expectation of the model's accuracy. The light-meson 
electromagnetic form factors were calculated in impulse 
approximation~\protect\cite{cdrpion,Maris:1998hc,thomson} and $\xi(w)$ was 
obtained from $f_+^{B\to D}(t)$ via Eq.~(\protect\ref{xif}).  The values in 
the ``Obs.''  column were taken from Refs.\ 
\protect\cite{mr97,pdg98,Amendolia:1986wj,flynn,richman,cesr96,latt,gHsHpi}. 
(Adapted from Ref.\ \protect\cite{mishaSVY}.)} 
\begin{center} 
\renewcommand{\arraystretch}{1.4} 
\setlength\tabcolsep{5pt} 
\begin{tabular}{lll|lll} 
\hline\noalign{\smallskip} 
        & Obs.  & Calc. & & Obs.  & Calc. \\\hline 
$f_\pi$   & 0.131 & 0.146 & $m_\pi$   & 0.138 & 0.130 \\ 
$f_K  $   & 0.160 & 0.178 & $m_K$     & 0.496 & 0.449 \\ 
$\langle \bar u u\rangle^{1/3}$ & 0.241 & 0.220 & 
        $\langle \bar s s\rangle^{1/3}$ & 0.227 & 0.199\\ 
$\langle \bar q q\rangle_\pi^{1/3}$ & 0.245 & 0.255& 
        $\langle \bar q q\rangle_K^{1/3}$ & 0.287 & 0.296\\ 
$f_\rho$   & 0.216      & 0.163 & 
        $f_{K^\ast}$   & 0.244      & 0.253 \\ 
$\Gamma_{\rho\pi\pi}$ & 0.151     & 0.118 & 
        $\Gamma_{K^\ast (K\pi)}$  & 0.051     & 0.052 \\ 
$f_D$   & 0.200 $\pm$ 0.030     & 0.213 & 
        $f_{D_s}$ & 0.251 $\pm$ 0.030     & 0.234 \\ 
$f_B$   & 0.170 $\pm$ 0.035      & 0.182 & 
$g_{B K^\ast \gamma} \hat M_b$ & 2.03 $\pm$ 0.62 & 2.86 \\\hline 
$f_+^{B\to D}(0)$ & 0.73 & 0.58  & 
        $f_\pi r_\pi$ & 0.44 $\pm$ 0.004 & 0.44   \\ 
$F_{\pi\,(3.3\,{\rm GeV}^2)}$ & 0.097 $\pm$  0.019  & 0.077 & 
        B$(B\to D^\ast)$ & 0.0453 $\pm$ 0.0032 & 0.052\\ 
$\rho^2$ &  1.53 $\pm$ 0.36 & 1.84 & 
        $\alpha^{B\to D^\ast}$ & 1.25 $\pm$ 0.26 & 0.94 \\ 
$\xi(1.1)$  & 0.86 $\pm$ 0.03& 0.84 & 
        $A_{\rm FB}^{B\to D^\ast}$ & 0.19 $\pm$ 0.031 & 0.24 \\ 
$\xi(1.2)$  & 0.75 $\pm$ 0.05& 0.72 & 
        B$(B\to \pi)$ & (1.8 $\pm$ 0.6)$_{\times 10^{-4}}$  & 2.2 \\ 
$\xi(1.3)$  & 0.66 $\pm$ 0.06& 0.63 & 
        $f^{B\to \pi}_{+\,(14.9\,{\rm GeV}^2)}$ & 0.82 $\pm$ 0.17 & 0.82 \\ 
$\xi(1.4) $ & 0.59 $\pm$ 0.07& 0.56 & 
        $f^{B\to \pi}_{+\,(17.9\,{\rm GeV}^2)}$ & 1.19 $\pm$ 0.28 & 1.00 \\ 
$\xi(1.5) $ & 0.53 $\pm$ 0.08& 0.50 & 
        $f^{B\to \pi}_{+\,(20.9\,{\rm GeV}^2)}$ & 1.89 $\pm$ 0.53 & 1.28 \\ 
B$(B\to D)$ & 0.020 $\pm$ 0.007 & 0.013 & 
        B$(B\to \rho)$ & (2.5 $\pm$ 0.9)$_{\times 10^{-4}}$ & 4.8 \\ 
B$(D\to K^\ast)$ & 0.047 $\pm$ 0.004  & 0.049 & 
        $f_+^{D\to K}(0)$ & 0.73 &  0.61 \\ 
$\displaystyle\frac{V(0)}{A_1(0)}^{D \to K^\ast}$ & 1.89 $\pm$ 0.25 & 1.74 & 
        $f_+^{D\to \pi}(0)$ & 0.73 &  0.67 \\ 
$\displaystyle\frac{\Gamma_L}{\Gamma_T}^{D \to K^\ast}$ & 1.23 $\pm$ 0.13 & 1.17 & 
        $g_{B^\ast B\pi}$ & 23.0 $\pm$ 5.0 & 23.2 \\ 
$\displaystyle\frac{A_2(0)}{A_1(0)}^{D \to K^\ast}$ & 0.73 $\pm$ 0.15 & 0.87 & 
        $g_{D^\ast D\pi}$ & 10.0 $\pm$ 1.3 & 11.0 \\\hline 
\end{tabular} 
\end{center} 
\end{table} 
 
It is evident that the fitted heavy-quark masses are consistent with the 
estimates in Ref.\ \cite{pdg98} and hence that the heavy-meson binding energy 
is large: 
\begin{equation} 
E_D:= m_D - \hat M_c = 0.67\,{\rm GeV}\,,\; 
E_B:= m_B - \hat M_b = 0.70\,{\rm GeV}\,. 
\end{equation} 
These values yield $E_D/\hat M_c= 0.51$ and $E_B/\hat M_b= 0.15$, which
furnishes another indication that while an heavy-quark expansion is accurate
for the $b$-quark it will provide a poor approximation for the $c$-quark.
This is emphasised by the value of $\omega_D = \omega_B$, which means that
the Compton wavelength of the $c$-quark is greater than the length-scale
characterising the bound state's extent.
 
\begin{figure}[t] 
\centerline{\includegraphics[width=.60\textwidth]{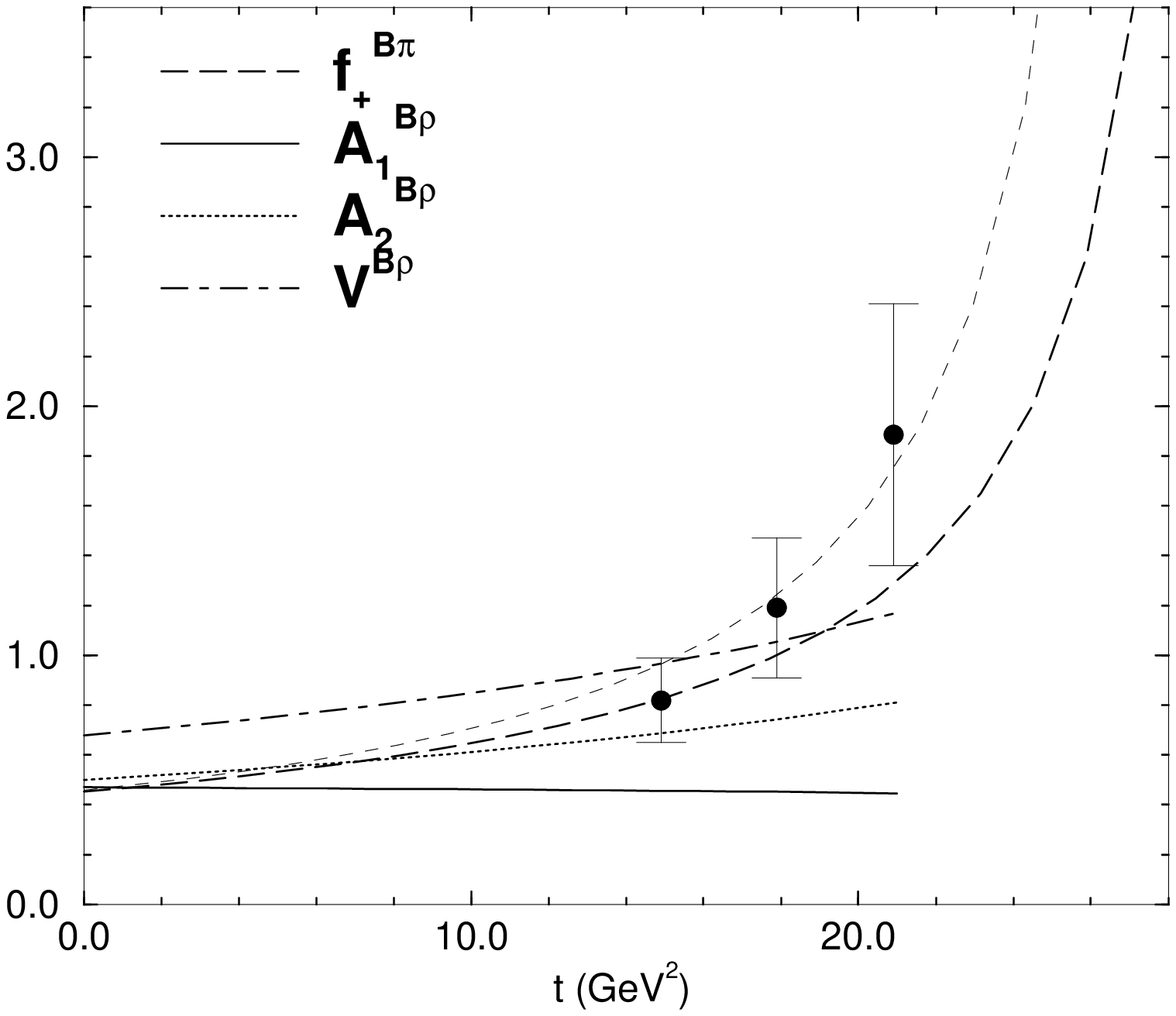}} 
\caption{\label{btopi} Semileptonic $B\to \pi$ and $B\to \rho$ form factors 
with, for comparison, data from a lattice simulation \protect\cite{latt} and 
a vector dominance, monopole model: $f_+^{B\to \pi}(t)= 
0.46/(1-t/m_{B^\ast}^2)$, $m_{B^\ast} = 5.325\,$GeV (\textit{short-dashed 
line}).  (Adapted from Ref.\ \protect\cite{mishaSVY}.)} 
\end{figure}

With the parameters fixed, in Ref.\ \cite{mishaSVY} values for a wide range
of other observables were calculated with the vast majority of the results
being true predictions.  The breadth of application is illustrated in Table
\ref{mishasummary}, and in Fig.\ \ref{btopi} which depicts the calculated
$t$-dependence of $B \to \pi\,,\rho$ semileptonic transition form factors.
I note that now there is a first experimental result for the $D^{\ast+}$
width \protect\cite{newDstar}: $\Gamma_{D^{\ast +}}= (96 \pm 4 \pm 22
)\,$keV, $g_{D^\ast D \pi} = 17.9 \pm 0.3 \pm 1.9$.  Its confirmation and the
gathering of additional information on $c$-quark mesons is crucial to
improving our knowledge of the evolution from the light- to the heavy-quark
domain, a transition whose true understanding will significantly enhance our
grasp of nonperturbative dynamics.
 
\begin{table}[t] 
\caption{\label{mishasummary} Predictions in Ref.\ \protect\cite{mishaSVY} 
for a selection of observables.  The ``Obs.'' values are extracted from 
Refs.~\protect\cite{pdg98,flynn,richman,newDstar,kaondat}. $t_{\rm max}$ is 
the maximum momentum transfer available in the process identified and 
$\omega_{\max} = \omega(t_{\rm max})$ calculated from Eq.\ 
(\protect\ref{defomega}). (Adapted from Ref.\ \protect\cite{mishaSVY}.)} 
\begin{center} 
\renewcommand{\arraystretch}{1.4} 
\setlength\tabcolsep{5pt} 
\begin{tabular}{lll|lll} 
\hline\noalign{\smallskip} 
        & Obs.  & Calc. & & Obs.  & Calc. \\\hline 
$f_K r_K$       &   0.472 $\pm$ 0.038 & 0.46 & 
        $-f_K^2 r_{K^0}^2$ &  (0.19 $\pm$ 0.05)$^2$ & (0.10)$^2$   \\ 
$g_{\rho\pi\pi}$ & 6.05 $\pm$ 0.02 & 5.27  & 
        $\Gamma_{D^{\ast 0} }$ (MeV)& $ < 2.1 $  & 0.020   \\ 
$g_{K^\ast K \pi^0}$ & 6.41 $\pm$ 0.06 & 5.96  & 
        $\Gamma_{D^{\ast +}}$ (keV) &  $96\pm4\pm 22$ & 37.9 \\ 
$g_\rho$ & 5.03 $\pm$ 0.012 & 5.27  & 
        $\Gamma_{D_s^{\ast} D_s \gamma}$ (MeV)& $< 1.9$  & 0.001    \\ 
$f_{D^\ast}$ (GeV) &   & 0.290   & 
        $\Gamma_{B^{\ast +} B^+ \gamma}$ (keV)&   & 0.030    \\ 
$f_{D^\ast_s}$ (GeV)&   & 0.298   & 
        $\Gamma_{B^{\ast 0} B^0 \gamma}$ (keV)&  &  0.015 \\ 
$f_{B_s}$ (GeV) & 0.195 $\pm$ 0.035 & 0.194  & 
        $\Gamma_{B_s^{\ast} B_s \gamma}$ (keV)&  &  0.011 \\ 
$f_{B^\ast}$ (GeV)&   & 0.200   & 
        B($D^{\ast +}\!\to D^+ \pi^0$) & 0.306 $\pm$ 0.025 & 0.316 \\ 
$f_{B^\ast_s}$ (GeV)&   & 0.209   & 
        B($D^{\ast +}\!\to D^0 \pi^+$) & 0.683 $\pm$ 0.014 & 0.683 \\ 
$f_{D_s}/f_D$ & 1.10 $\pm$ 0.06 &  1.10  & 
        B($D^{\ast +}\!\to D^+ \gamma$) & 0.011~$^{+ 0.021}_{-0.007}$ & 0.001 \\ 
$f_{B_s}/f_B$  & 1.14 $\pm$ 0.08  & 1.07   & 
        B($D^{\ast 0}\!\to D^0 \pi^0$) &  0.619 $\pm$ 0.029 & 0.826 \\ 
$f_{D^\ast}/f_D$ &       &  1.36  & 
        B($D^{\ast 0}\!\to D^0 \gamma$) &  0.381 $\pm$ 0.029 &  0.174 \\ 
$f_{B^\ast}/f_B$  &       & 1.10   & 
        B($B \to K^\ast \gamma$) & (5.7 $\pm$ 3.3)$_{10^{-5}}$ & 11.4 \\ 
$R_1^{B\to D^\ast}(1)$ &   1.30 $\pm$ 0.39   & 1.32 & 
        $R_2^{B\to D^\ast}(1)$ & 0.64 $\pm$ 0.29 & 1.04   \\ 
$R_1^{B\to D^\ast}(w_{\rm max})$ &      & 1.23 & 
        $R_2^{B\to D^\ast}(w_{\rm max})$ &  & 0.98   \\ 
B($D^+\to \rho^0$) &   & 0.032   & 
        $\alpha^{D \to \rho}$ &   & 1.03  \\ 
${\rm B}(D^0\to K^-)$   &  0.037 $\pm$ 0.002  &  0.036   & 
        $\displaystyle\frac{{\rm B}(D\to \rho^0)} 
        {{\rm B}(D\to K^\ast)}$ & 0.044 $\pm$ 0.034  & 0.065   \\ 
$A_1^{D\to K^\ast}(0)$ & 0.56 $\pm$ 0.04 & 0.46   & 
        $A_1^{D\to K^\ast}(t_{\rm max}^{D\to K^\ast})$ &0.66 $\pm$ 0.05 & 0.47 \\ 
$A_2^{D\to K^\ast}(0)$ & 0.39 $\pm$ 0.08 & 0.40   & 
      $A_2^{D\to K^\ast}(t_{\rm max}^{D\to K^\ast})$ & 0.46 $\pm$ 0.09 & 0.44 \\ 
$V^{D\to K^\ast}(0)$ & 1.1 $\pm$ 0.2 & 0.80   & 
        $V^{D\to K^\ast}(t_{\rm max}^{D\to K^\ast})$ & 1.4 $\pm$ 0.3 & 0.92 \\ 
$\displaystyle\frac{{\rm B}(D^0\to \pi)} 
        {{\rm B}(D^0\to K)}$ & 0.103 $\pm$ 0.039  & 0.098   & 
        $f_+^{D\to K}(t_{\rm max}^{D\to K})$ & 1.31 $\pm$ 0.04 & 1.11 \\ 
$\displaystyle\frac{f_+^{D\to \pi}(0)}{f_+^{D\to K}(0)}$ & 1.2 $\pm$ 0.3 &  1.10  & 
        $f_+^{D\to \pi}(t_{\rm max}^{D\to \pi})$ &  & 2.18 \\ 
$R_1^{D\to K^\ast}(1)$ &      & 1.72 & 
        $R_1^{D\to K^\ast}(w_{\rm max})$ &      & 1.74 \\ 
$R_1^{D\to \rho}(1)$ &      & 2.08 & 
        $R_1^{D\to \rho}(w_{\rm max})$ &      & 2.03 \\\hline 
\end{tabular} 
\end{center} 
\end{table} 
 
\subsubsection{Fidelity of heavy-quark symmetry.} 
\begin{figure}[t] 
\centerline{\includegraphics[width=.60\textwidth]{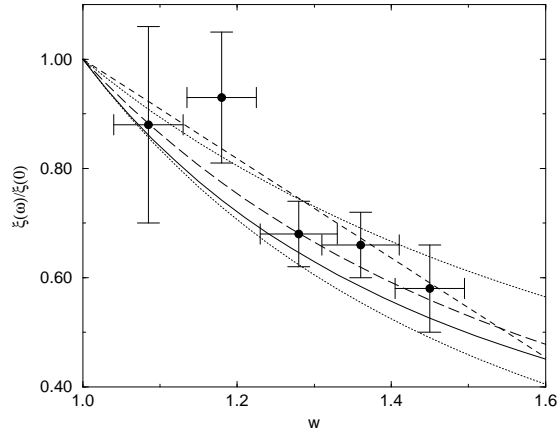}} \caption{\label{xiw} 
Calculated \protect\cite{mishaSVY} form of $\xi(w)$ (\textit{solid line}) 
compared with experimental analyses: linear fit from Ref.\ 
\protect\cite{cesr96}, Eq.~(\protect\ref{cesrlinear}) (\textit{short-dashed 
line}); nonlinear fit from Ref.\ \protect\cite{cesr96}, 
Eq.~(\protect\ref{cesrnonlinear}) (\textit{long-dashed line}).  The two lighter 
dotted lines are the nonlinear fit of Ref.\ \protect\cite{cesr96} evaluated 
with the extreme values of $\rho^2$: upper line, $\rho^2= 1.17$ and lower line, 
$\rho^2=1.89$.  The data points are from Ref.\ \protect\cite{argus93}. (Adapted 
from Ref.\ \protect\cite{mishaSVY}.)} 
\end{figure}

The universal function characterising semileptonic transitions in the
heavy-quark symmetry limit, $\xi(w)$ introduced in Sec.\ \ref{sec:HQSym}, can
be estimated most reliably from $B\to D,D^\ast$ transitions.  Using Eq.\
(\ref{xif}) to infer this function from $f_+^{B\to D}(t)$, one obtains
\begin{equation} 
\label{xifp} 
\xi^{f_+}(1)= 1.08\,, 
\end{equation} 
which is a measurable deviation from Eq.\ (\ref{xione}).  The ratio 
$\xi^{f_+}(w)/\xi^{f_+}(0)$ is depicted in Fig.\ \ref{xiw}, wherein it is 
compared with two experimental fits \cite{cesr96}: 
\begin{eqnarray} 
\xi(w) & = & 1 - \rho^2\,( w - 1), \rule{6.8em}{0ex} 
\rho^2 = 0.91\pm 0.15 \pm 0.16\,, 
\label{cesrlinear}\\ 
\xi(w) & = & \frac{2}{w+1}\,\exp\left[(1-2\rho^2) \frac{w-1}{w+1}\right], 
        \;\rho^2 = 1.53 \pm 0.36 \pm 0.14\,. 
\label{cesrnonlinear} 
\end{eqnarray} 
The evident agreement was possible because Ref.\ \cite{mishaSVY} did not 
employ the heavy-quark expansion of Eq.\ (\ref{hqf}), in particular and 
especially not for the $c$-quark.  The calculated result (\textit{solid 
curve}) in Fig.\ \ref{xiw} is well approximated by \begin{equation} 
\label{fitIW} 
\xi^{f_+}(w) = \frac{1}{1 + \tilde\rho^2_{f_+}\,(w-1)  }\,, 
\;\tilde\rho^2_{f_+}=1.98\,. 
\end{equation} 
 
Equations (\ref{a1xi}) were also used in Ref.\ \cite{mishaSVY} to extract 
$\xi(w)$ from $B\to D^\ast$.  This gave $\xi^{A_1}(1)= 0.987$, 
$\xi^{A_2}(1)=1.03$, $\xi^{V}(1)= 1.30$, an $w$-dependence well-described by 
Eq.\ (\ref{fitIW}) but with $\tilde\rho^2_{A_1}=1.79$, 
$\tilde\rho^2_{A_2}=1.99$, $\tilde\rho^2_{V}=2.02$, and the ratios, Eqs.\ 
(\ref{Rratios}), $R_{1}(1)/R_{1}(w_{\rm max})=1.08$, $R_{2}(1)/R_{2}(w_{\rm 
max})=1.06$. 
 
This collection of results indicates the degree to which heavy-quark 
symmetry is respected in $b\to c$ processes.  Combining them it is clear that 
even in this case, which is the nearest contemporary realisation of the 
heavy-quark symmetry limit, corrections of $\lsim 30$\% must be expected.  In 
$c\to s\,,d$ transitions the corrections can be as large as a factor of two, 
as evident in Table \ref{mishasummary}. 
 
\section{Epilogue} 
\label{sec:epilogue} 
This contribution provides a perspective on the modern application of 
Dyson-Schwin\-ger equations (DSEs) to light- and heavy-meson properties.  The 
keystone of this approach's success is an appreciation and expression of the 
mo\-men\-tum-dependence of dressed-parton propagators at infrared 
length-scales.  That dependence is responsible for the magnitude of 
constituent-quark and -gluon masses, and the length-scale characterising 
confinement in bound states; and is now recognised as a fact. 
 
It has recently become clear that the simple rainbow-ladder DSE truncation is 
the first term in a systematic and nonperturbative scheme that preserves the 
Ward-Takahashi identities which express conservation laws at an hadronic level. 
This has enabled the proof of exact results in QCD, and explains why the 
truncation has been successful for light vector and flavour nonsinglet 
pseudoscalar mesons. Emulating more of these achievements with \textit{ab 
initio} calculations of heavy-meson properties is a modern challenge. 
 
However, at present, the study of heavy-meson systems using DSE methods
stands approximately at the point occupied by those of light-meson properties
seven -- eight years ago.  A Poincar\'e covariant treatment exploiting
essential features, such as propagator dressing and sensible bound state
Bethe-Salpeter amplitudes, has been shown capable of providing a unified and
successful description of light- and heavy-meson observables.  The goal now
is to make the case compelling by tying the separate elements together;
namely, relating the propagators and Bethe-Salpeter amplitudes via a single
kernel.  I am confident this will be accomplished, and the DSEs become a
quantitatively reliable and intuition building tool as much in the
heavy-quark sector as they are for light-quark systems.
 
While a more detailed understanding will be attained in pursuing this goal,
certain qualitative results established already are unlikely to change.  For
example, it is plain that light- and heavy-mesons are essentially the same,
they are simply bound states of dressed-quarks.  Moreover, the magnitude of
the $b$-quark's current-mass is large enough to sustain heavy-quark
approximations for its propagator and the amplitudes for bound states of
which it is a constituent.  In addition, and unfortunately in so far as
practical constraints on the Standard Model are concerned, the current-mass
of the $c$-quark is too small to validate an heavy-quark approximation.
 
\section*{Acknowledgments} 
I am grateful for the hospitality and support of my colleagues and the staff 
in the Bogoliubov Laboratory of Theoretical Physics at the Joint Institute 
for Nuclear Research, Dubna, Russia. 
This work was supported by: the Department of Energy, Nuclear Physics 
Division, under contract no.\ \mbox{W-31-109-ENG-38}; Deutsche 
For\-schungs\-ge\-mein\-schaft, under contract no.\ Ro~1146/3-1; and 
benefited from the resources of the National Energy Research Scientific 
Computing Center.

%
 

\begin{thebibliography}{99.}
\bibitem{fn:foot} The fitting used \cite{pdg98}: $V_{ub}=0.0033$, $V_{cd}= 
0.2205$, $V_{cs}= 0.9745$ and $V_{cb}= 0.039$; and, in GeV, $M_\rho= 0.77$, 
$M_{K^\ast}=0.892$ and, except in the kinematic factor 
$\lambda(m_1^2,m_2^2,t)$ where the splittings are crucial, averaged $D$- and 
$B$-meson masses: $m_D= 1.99$, $m_B=5.35$ (from $m_D=1.87$, $m_{D_s}=1.97$, 
$M_{D^\ast}=2.01$, $M_{D_s^\ast}=2.11$, and $m_B=5.28$, $m_{B_s}=5.37$, 
$M_{B^\ast}=5.32$, $M_{B_s^\ast}=5.42$).  Furthermore, $b_0^{u}=0.131$, 
$b_0^{s}=0.105$, $b_3^{u}=0.185=b_3^{s}$ and $\epsilon = 10^{-4}$ were not 
varied, being instead fixed at the values determined in Ref.\ 
\protect\cite{thomson}.
\end{thebibliography}
\end{document}